\documentclass[acmsmall]{acmart}

\usepackage{booktabs}
\usepackage{subcaption, multirow}
\usepackage{color, colortbl}
\usepackage{graphicx}
\usepackage{longtable}
\usepackage{pifont}
\usepackage{titlesec}

\AtBeginDocument{%
  \providecommand\BibTeX{{%
    \normalfont B\kern-0.5em{\scshape i\kern-0.25em b}\kern-0.8em\TeX}}}

\setcopyright{rightsretained}
\acmJournal{PACMHCI}
\acmYear{2022} \acmVolume{6} \acmNumber{CSCW2} \acmArticle{341} \acmMonth{11} \acmPrice{}\acmDOI{10.1145/3555761}

\begin{document}

\title{Understanding Questions that Arise When Working with Business Documents}


\author{Farnaz Jahanbakhsh}
\authornote{Research performed while interning at Microsoft Research.}
\affiliation{%
  \institution{Computer Science and Artificial Intelligence Laboratory, Massachusetts Institute of Technology}
  \city{Cambridge}
  \country{USA}
}
\email{farnazj@mit.edu}

\author{Elnaz Nouri}
\affiliation{%
  \institution{Microsoft Research}
  \city{Redmond}
  \country{USA}
}
\email{Elnaz.Nouri@microsoft.com}

\author{Robert Sim}
\affiliation{%
  \institution{Microsoft Research}
  \city{Redmond}
  \country{USA}
}
\email{rsim@microsoft.com}

\author{Ryen W. White}
\affiliation{%
  \institution{Microsoft Research}
  \city{Redmond}
  \country{USA}
}
\email{ryenw@microsoft.com}

\author{Adam Fourney}
\affiliation{%
  \institution{Microsoft Research}
  \city{Redmond}
  \country{USA}
}
\email{Adam.Fourney@microsoft.com}

\renewcommand{\shortauthors}{Farnaz Jahanbakhsh et al.}

\begin{abstract}
 While digital assistants are increasingly used to help with various productivity tasks, less attention has been paid to employing them in the domain of business documents. To build an agent that can handle users' information needs in this domain, we must first understand the types of assistance that users desire when working on their documents. In this work, we present results from two user studies that characterize the information needs and queries of authors, reviewers, and readers of business documents. In the first study, we used experience sampling to collect users’ questions in-situ as they were working with their documents, and in the second, we built a human-in-the-loop document Q\&A system which rendered assistance with a variety of users' questions. Our results have implications for the design of document assistants that complement AI with human intelligence including whether particular skillsets or roles within the document are needed from human respondents, as well as the challenges around such systems.

\end{abstract}



\maketitle

\section{Introduction}

Systems and software are increasingly incorporating intelligent digital assistants in order to help people be ever more productive even as systems, documents, and information spaces become more complex. For example, at home, people can rely on their voice assistants to manage processes such as shopping, cooking, and to quickly retrieve information from the web \cite{IFTT2017Forbes, Cooking2017GoogleHome, hoy2018alexa}. At work, digital assistants help with scheduling meetings, triaging emails, and managing task lists \cite{cranshaw2017calendar, dredze2009intelligent, myers2007intelligent}.

One workplace area of interest that has not seen significant progress is document-centered assistance. In this scenario, people engage with an intelligent digital assistant to consume and operate over written documents to perform complex tasks more quickly. Such assistance can also be useful in contexts where working with business documents is challenging, for example, when using a mobile phone ~\cite{iqbal2018multitasking}.

To achieve this vision and build an agent that is prepared to handle a wide variety of requests from the user, we must first understand and characterize the types of assistance that people would want as they are working on their documents. While prior work has studied the types of queries that people would generate given one or a collection of public documents \cite{dunn2017searchqa, kwiatkowski2019natural, rajpurkar2016squad, yang2015wikiqa}, it is conceivable that these queries may not generalize to private or business documents that are in various stages of preparation. The information needs in a document-centric context could be especially different given that users may have prior interactions with a document and may not want to only read, but also review, or add information to the document. Therefore, to understand users' actual needs, it is important that we involve authors, reviewers, and readers of such documents in the process of generating queries. In addition, for these questions to have ecological validity, we must collect them in-situ as users are working on their documents and their information needs arise.


In this work, we conducted two user studies to gain insight into the distribution of queries with which users need support in a document-centric scenario. Our focus in these studies was on answering questions rather than on executing commands or taking direct action. For example, participants could ask about facts, or for technical instruction, but were discouraged from making requests for the system to directly alter document content.
In the first study, we collected questions via an experience sampling method~\cite{larson2014experience} by having participants install a Microsoft Word add-in which, at random points in time while they were working on a document, would prompt them with a short questionnaire asking about their current information need, but did not directly answer participant queries. In the second study, participants submitted their questions via an add-in and received answers from a human-in-the-loop document Q\&A system. Given a document and a question, a Q\&A ML model extracted a passage out of the document as a candidate answer \cite{ter2020conversations}. A human worker would then decide between transmitting the candidate answer to the study participant, or composing and transmitting their own answer in cases where the AI-provided answer was incorrect or insufficient.

 We expected that answering some of the potential questions would be beyond the capabilities of current systems and there may be a need for human intelligence. We hypothesized that if a Q\&A system in this domain were to be deployed to accommodate user needs, at least for the present, it would need to incorporate human workers. The studies would then allow us to investigate which types of questions needed human worker support, to what extent this support was needed, and what capabilities the human workers should have to accommodate the questions. We conducted the second study as a technology probe into the problem space, to not only understand the use of a document Q\&A assistant in a real world setting, but also field test the technology that is available, and understand how we can address the inadequacies of the current technology~\cite{hutchinson2003technology}. 

The settings of both of these studies, the first simply asking about information needs and the second augmenting the AI with human intelligence, allowed users to be liberal in the questions they posed and not limited to the capabilities of current digital assistants, or the in-application search capabilities integrated in some productivity software (e.g., \cite{bota2018characterizing}). Therefore, these approaches allow us to investigate what capabilities to incorporate in document assistants prior to making deep investments in their development. We found for instance, that while factual questions, i.e., questions that can be answered by extracting a passage out of a document, are the focus of state-of-the-art Q\&A datasets and models \cite{nguyen2016ms, rajpurkar2016squad}, these types of questions are in fact not asked often in the domain of business documents. In addition, the study results give us an understanding that, for the foreseeable future, effective assistance will likely involve at least some human-in-the loop or hybrid intelligence---important classes of questions are best answered by document authors, collaborators, domain experts, and other types of human respondents. Even in these cases, AI may assist in question triage, and in routing questions to the appropriate parties. 

The main contributions of this work are the characterization of the types of assistance that users desire when working on business documents and implications for the design of digital assistants that can provide such assistance. Specifically, our research seeks to answer the following research questions: 

\begin{itemize}
    \item What types of document-related questions do people desire support with from a digital assistant?
   \item How are the distribution of questions different depending on whether the user is an author, reviewer, or reader of the document?
   \item What skillsets and roles are needed from human respondents to accommodate requests in a human-in-the-loop document Q\&A assistant?

\end{itemize}

\section{Related Work}
We situate our work in the context of prior work on document question answering, automated and hybrid digital assistants, and studies of tools that facilitate authoring, reviewing, or reading documents.

\subsection{Document Question Answering}
Recent years have seen significant progress in AI based methods for finding answers to questions given one or a collection of documents. This body of work includes factoid Q\&A, document understanding, summarization, and comparison \cite{rajpurkar2018know, iyyer2014neural, chen2017reading, yao2017recent, xiao2018dual, han2020learning, gao2017entity, gelbukh1999document}.

Most datasets for document question answering contain factoid questions generated by crowd workers or search queries on documents found on the Web \cite{dunn2017searchqa, joshi2017triviaqa, kwiatkowski2019natural, nguyen2016ms, yang2018hotpotqa, rajpurkar2016squad, yang2015wikiqa}. Models trained on these datasets therefore may fail to generalize to personal and business documents with which people have richer context and possibly prior interactions. Indeed, previous work in the context of email and Web search has shown that people’s information needs are different depending on whether they are a co-owner of a document \cite{ai2017characterizing}. 

Closer to our scenario, Ter Hoeve et al. leveraged crowd workers, and a corpus of public documents mined from the Web, to investigate the conversational assistance that people would want in a document consumption scenario. The workers were trained to imagine having some familiarity with the document subject matter, and to produce questions about a document based on its summary~\cite{ter2020conversations}. However, all documents were presented in their final published forms, and participants often assumed the role of ``reader'', with only limited knowledge of document content or provenance. Related is a study by Todi et al. that explored the types of information users would want to access conversationally from a GUI dataset. The researchers asked a set of designers, developers and end users, with no prior knowledge about the kinds of GUIs that could be found in the dataset, to pose queries to a hypothetical chatbot that would help seek information from the dataset~\cite{todi2021conversations}.

We extend the body of work on question answering to better understand people's information needs while working with business documents across various states of preparation and familiarity. 
To this end, it is imperative that we consider queries from authors, reviewers, and readers of such documents, and capture their actual needs which may be unique to their role and context. It is also desirable for these questions to be captured in-situ as users' needs arise. Therefore, in this work, we use two methods: experience sampling and a human-in-the-loop document Q\&A system, to capture users' information needs while they are working on their documents. 

\subsection{Digital and Hybrid Assistance}

Digital agents are increasingly used in areas such as entertainment, e-commerce, healthcare, and to help with various productivity tasks \cite{hoy2018alexa, ferguson2009cardiac, olson2019voice}. 
In the workplace, such assistants help with
scheduling meetings, triaging emails, managing task lists, and even performing data science tasks \cite{cranshaw2017calendar, myers2007intelligent, dredze2009intelligent, fast2018iris}.  In the context of documents, digital assistants now perform rich pre-defined tasks, such as PowerPoint theme suggestions, or intelligent placeholders in Microsoft Word \cite{powerpoint, Word2018todo}. Document-centric assistance can also be useful in contexts that have been reported as challenging, e.g., when using a mobile phone~\cite{iqbal2018multitasking}, or when in a vehicle~\cite{martelaro2019}.

Although these automated systems can accomplish many tasks, they still have limited capabilities in understanding complex or nuanced requests, or when requests simply fall outside of the systems' domains \cite{tur2014detecting}. 
To augment their abilities, a body of work has attempted to incorporate human intelligence into the workflow of such systems \cite{kamar2016hybrid}. For instance, Lasecki et al. developed Chorus, a conversational agent that allows users to interact with a group of crowd workers as if they are a single conversational partner \cite{lasecki2013chorus}. Commercial services such as Facebook M, Clara, and X.ai also employ a hybrid of machine and human knowledge to run errands for consumers \cite{hempel2015facebook, huet2016humans, cade2015wired}.

Closest to this scenario and to our work is Soylent, a word processor add-in that employs crowdworkers to help users edit Word documents \cite{bernstein2010soylent}. Indeed, Soylent serves as an inspirational model for our work, but is limited in that it considers only a few specific editing scenarios, and largely overlooks opportunities to facilitate document consumption. In consumption scenarios, it is conceivable that different questions should be routed to different types of respondents, similar to social Q\&A systems such as \textit{IM-an-Expert}~\cite{richardson2011supporting}, \textit{Aardvark}~\cite{horowitz2010anatomy}, and \textit{Zephyr}~\cite{ackerman1996zephyr} which routed questions to individuals with expertise or interest in the subject matter of the question.

Our work aims to bridge these gaps by examining the types of document-centric questions with which people need support. It is conceivable that for the foreseeable future, accommodating some requests in this domain requires some degree of human assistance. Understanding users' needs, including what types of human intelligence we should incorporate into such a system is the first step towards designing systems that can accommodate these needs.

\subsection{Tools that Support Authoring, Reviewing, or Reading Documents}

A body of work has developed or studied the usage of tools that help with authoring, reviewing, or reading documents. 
Among those that studied the usage of such tools, some characterized how people use existing collaborative document authoring tools such as Google Docs and Microsoft OneDrive~\cite{olson2017people, yim2017synchronous, sun2014collaboration}. For instance, research reported on the role that the edit history feature plays in assessing a team's progress and the amount of individual effort~\cite{birnholtz2012tracking}. Birnholtz et al. found that users write utterances in the documents that are not related to the content but are instead intended 
for communicating with their collaborators~\cite{birnholtz2013write}. Wang et al. found that document collaborators often have different inherited role structures (e.g., employees and managers) and that current authorship tools do not support these roles. For instance, students and employees reported that they do not feel comfortable editing the text written by their advisors (or managers)~\cite{wang2017users}.
Posner and Baecker characterized the collaborative writing process as including six activities: brainstorm, research, plan, write, edit, and review. They suggested that a collaborative writing tool should support all these activities~\cite{posner1992people}.

A thread of work has focused on tools that help with authoring or reading of medical records. This domain has been given special attention because physicians often need to retrieve a complete picture from a patient’s records in just a few minutes before a consultation or emergency procedure~\cite{sultanum2018doccurate}. To facilitate reading and extracting information from electronic health records, a line of work has presented automated techniques for summarizing these records and generating timelines of patients’ problems~\cite{hallett2008multi, hallett2006summarisation}. Sultanum et al. presented Doccurate, a tool for visualizing large patient data records that allows physicians to curate concepts (e.g., cardiovascular issues) at the desired level of granularity for their practice~\cite{sultanum2018doccurate}.
Murray et al. developed MedKnowts to help facilitate the practice of reading medical data as well as inputting it. MedKnowts is a text editor for electronic health records that presents a unified interface to document symptoms, retrieve past records, medicines, lab results, etc. for better efficiency and error prevention. The tool enables automatic structured data capturing via a natural language interface and offers features such as autocomplete~\cite{Murray2021MedKnowts}.

In our work, we use the interaction paradigm of a digital assistant to facilitate task completion or accessing information in the domain of business documents. This framing would allow us to gain a comprehensive view of the types of queries with which authors, reviewers, and readers of documents need help; whereas framing the embodiment of interaction otherwise, such as a search engine or a tool with a simpler user interface, could have led users to constrain their questions for instance, to those that could be answered by document or passage retrieval.


\section{Method}

To investigate our three research questions, we designed a two-phase study. In the first phase, we collected document-oriented questions with which people reported needing support via an experience-sampling method. In this first phase, participants detailed their information needs, but did not expect, nor were provided with, answers to their queries. In the second phase, we designed and developed a human-in-the-loop system that collected users’ questions about their documents and provided responses to them. This system consisted of an AI component and human operators who supervised the AI and answered questions if the AI failed to produce a satisfactory answer.

The experience sampling in phase 1 was a step along the path to our vision of a hybrid human-AI document Q\&A system. Phase 1 served to inform our choices for the prototype and second study phase. Specifically, phase 1 provided us with initial insights into the types of questions people might ask, as well as the types of people who might need to be recruited to answer those questions -- a key factor in staffing a hybrid Q\&A system. Conversely, the tool we built in phase 2 was an actual prototype of the type of Q\&A system that we ultimately envision. The prototype directly enabled us to gain insight into the feasibility of generating answers in a document Q\&A context. It also allowed us to view the documents together with the questions, and obtain feedback about answer quality and system usefulness.

For the questions to have ecological validity, it was important that we collect them in-situ as users were authoring, reviewing, or reading a document. Therefore, we opted for using an add-in that would open on the side of a document, so that users could ask questions when they were working on a document and without leaving the application. For each phase of the study, we designed and developed an add-in that would work in Microsoft Word documents. 

As noted earlier, the scope of document assistance in both studies was question answering rather than application commanding and automation. This focus was conveyed through the wording of the prompts as well as in the recruitment emails and consent forms where we outlined that the study purpose is to understand what kinds of questions users have about documents they visit. In both phases, after consenting to the study, participants filled out a survey asking about demographics, their Word document consumption, and the readership of the Word documents they authored. Then they received instructions on how to install the add-in. Our study was approved by the Institutional Review Board in our organization.

\subsection{Phase 1 - Experience Sampling}

\begin{figure*}[!t]
    \centering
    \begin{minipage}{0.7\textwidth}
        \centering
        \includegraphics[width=\textwidth]{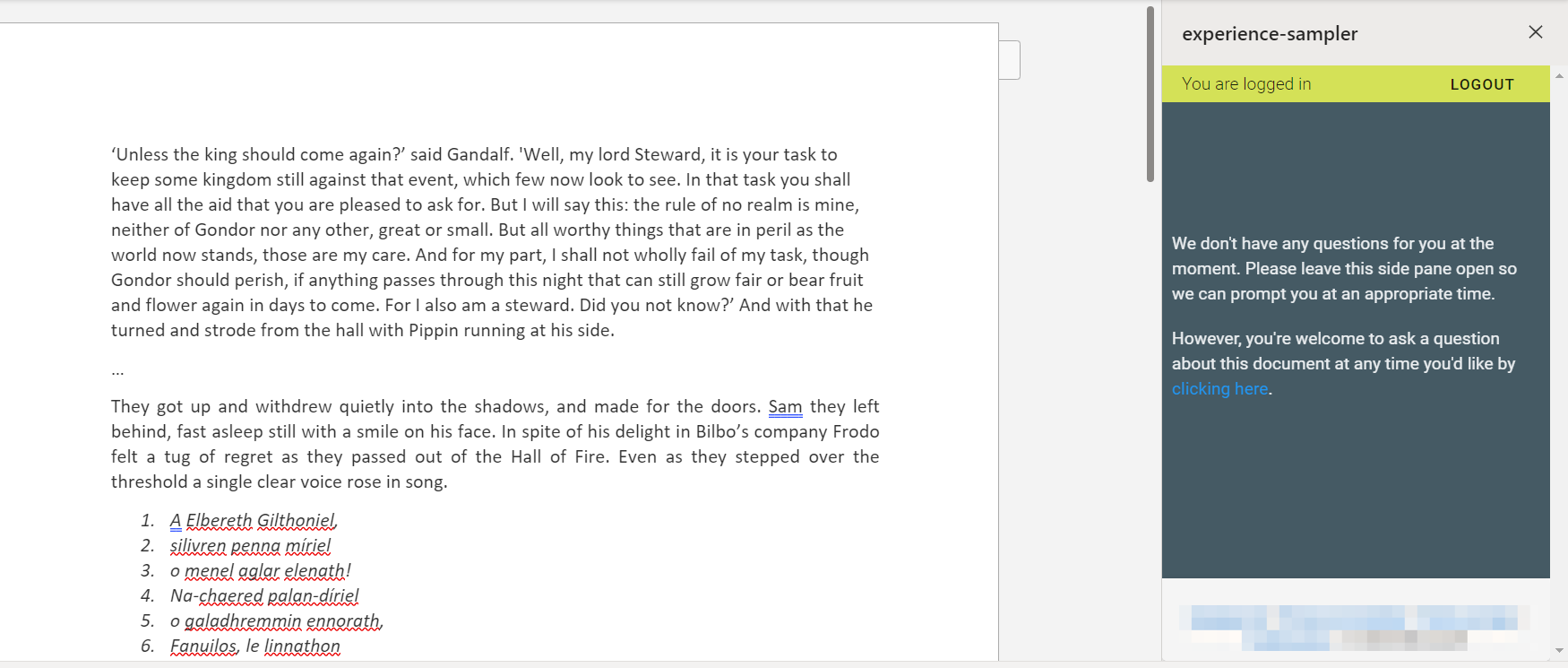} %
    \end{minipage}\hfill
    \begin{minipage}{0.3\textwidth}
        \centering
        \includegraphics[width=0.9\textwidth]{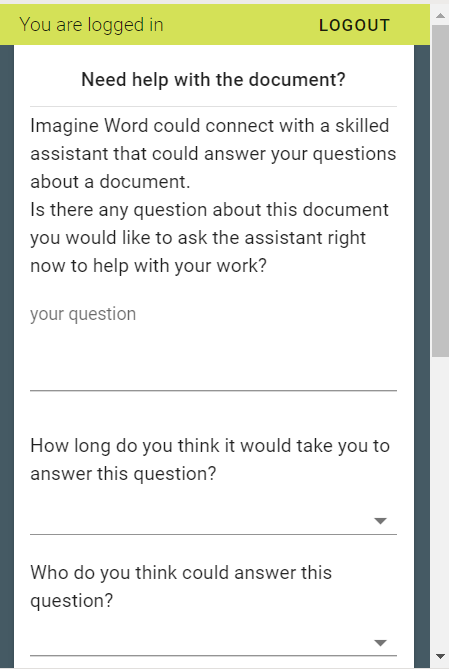} 
    \end{minipage}
    \caption{The figure on the left shows the Experience Sampler add-in opened in a side pane in a Word document. The screenshot is captured at a time before the user is prompted about their information need. The figure on the Right shows the questionnaire that the add-in presents to the user when the user is prompted.}
	\label{fig:phase_one}
\end{figure*}

At random points in time, the experience-sampler Microsoft Word add-in would prompt the user asking what question they had about the document at that moment, in addition to a few follow-up questions to better understand the context (Figure~\ref{fig:phase_one}).

To communicate the type of system we were envisioning, the first question was \textit{``Imagine Word could connect with a skilled assistant that could answer your questions about a document. Is there any question about this document you would like to ask the assistant right now to help with your work?''} To understand the perceived complexity of the question, the add-in then asked \textit{``How long do you think it would take you to answer this question?''} To gain insight into how questions can be routed to respondents in a human-in-the-loop Q\&A system, the add-in also asked \textit{``Who do you think could answer this question?''} The answer choices to this question included: \textit{``The author''}, \textit{``A domain expert''}, \textit{``A Microsoft Word expert''}, \textit{``Someone familiar with the doc''}, \textit{``Someone with enough time to read the doc''}, \textit{``Other''} along with a free-form text for elaboration, and \textit{``N/A''}. Because we expected the role of the user in the document to impact the questions that they ask, we also collected information about their contribution to the document: \textit{``What best describes your primary role in this document?''} (Reader, Reviewer, Author/Co-author), \textit{``Have you contributed to this document?''} (I have edited the document, I have commented on the document, I have not contributed to the document), \textit{``How long ago did you contribute to this document?''} (In the past 24 hrs, In the past week, Longer ago, Never).

In addition to capturing user responses, at the time of submitting the add-in survey, the add-in collected contextual information about the document including the document’s number of words, images, tables, lists, comments, timestamps (e.g., creation date, last modified date, date comments were posted, etc.), number of authors, filename, file size, URL, the location of the user’s cursor when asking the question, and whether the user submitting the survey was the creator, author, or the most recent contributor to the document. While we did not directly collect document content, we note that filenames often resembled  document titles\footnote{Microsoft Word uses the document title as the default filename when saving.}.

To minimize fatigue, we limited the number of prompts to six per day. Participants could answer the experience sampling questions whenever they wanted without waiting for the prompt. At the conclusion of the one-week period of the study, we had collected a total of 101 questions. Then we asked participants to open Word or their cloud documents homepage, select one or a few Word documents from their Recommended or Most Recently Opened lists of documents, and submit up to five questions for these documents as if they were working with the documents at that time. The completion of this step was voluntary and participants were not compensated additionally for submitting these questions. This step resulted in 38 more questions. We inspected these questions and found them to be extremely similar to those input earlier in the week, and thus we include them in our analyses. In total, questions from Phase 1 were asked about 103 distinct documents by 59 participants.


\subsection{Phase 2 - Human-in-the-loop Q\&A System}

\begin{figure*}[!t]
    \centering
    \begin{minipage}{0.33\textwidth}
        \centering
        \includegraphics[width=0.82\textwidth]{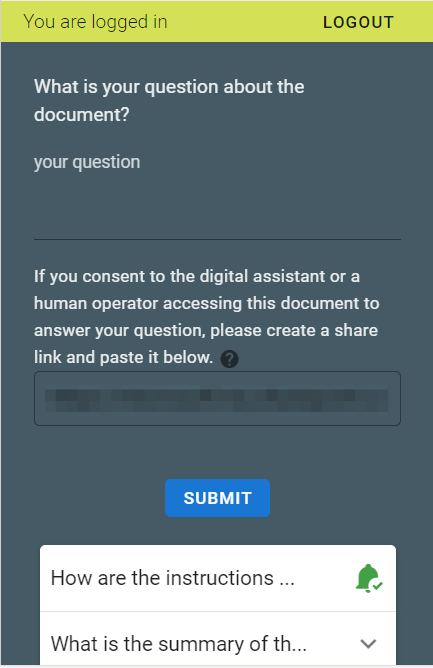} %
    \end{minipage}\hfill
    \begin{minipage}{0.33\textwidth}
        \centering
        \includegraphics[width=0.87\textwidth]{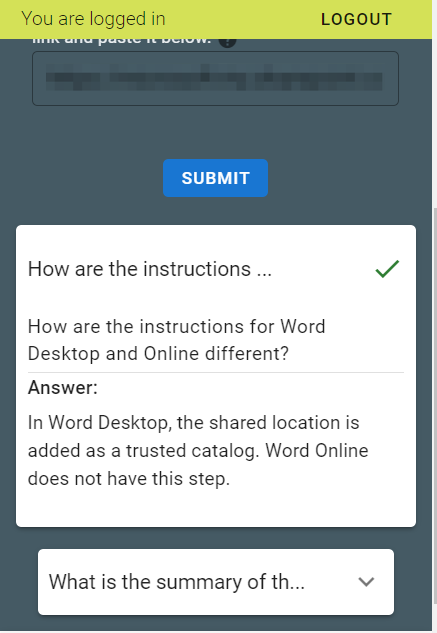} 
    \end{minipage}
    \begin{minipage}{0.33\textwidth}
        \centering
        \includegraphics[width=0.82\textwidth]{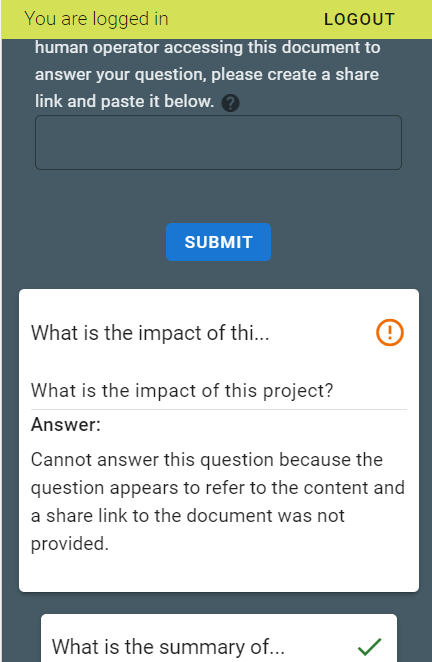}
    \end{minipage}
    \caption{The Q\&A add-in from Phase 2. The image on the left shows some questions that the user has asked about the document on the collapsed tiles. The green notification icon indicates that the question has been answered but the user has not yet viewed the answer. The image in the center displays the full question and the answer. Once an answer is viewed, the notification icon is changed to a checkmark. The image on the right shows another question from another document which could not be answered. The status is shown with a warning icon.}
	\label{fig:phase_two_addin}
\end{figure*}

We built a human-in-the-loop Q\&A prototype for connecting users with knowledge workers who would supervise and complement an AI trained to answer document-centered questions \cite{ter2020conversations}. The AI was a BERT Large model fine-tuned on the SQuAD2.0 \cite{rajpurkar2018know} and DQA datasets \cite{ter2020conversations}. Given a question and a document, the model would extract a passage out of the document if it detected an answer for the question.
Participants submitted and received answers to their questions through a Microsoft Word add-in. With each question, participants submitted a share link to the document if they consented to the digital assistant or a human knowledge worker accessing the document to answer their question. We informed the participants that for the purpose of the study, the knowledge workers were limited to the researchers involved in the project, and noted that the researchers were employed at the same technology company as the participants, thus minimizing concerns about confidentiality---indeed the document share links required corporate credentials to access. If a user did not submit a share link to their document, we did not have access to the document's content. When submitting a question, the add-in would collect the same contextual information about the document as in Phase 1. For each document, the add-in would display the user's previously asked questions in the form of expandable tiles. Because of the hybrid nature of the system, and the high level of human supervision, questions were not instantaneously answered. Participants were informed that their questions would be answered on a best effort basis and there may be a delay in the responses that they would receive from the system. Once a question had been answered or deemed unanswerable, the answer or the explanation for why the question could not be answered would appear below the question on the question tile. An icon on the question tile would signal whether the question was answered yet or marked as unanswerable, and whether the user had seen the answer or the explanation yet. In addition, users would receive email notifications about received answers. Figure~\ref{fig:phase_two_addin} displays different states of the Q\&A add-in.

\begin{figure}[!t]
    \centering
    \begin{minipage}{0.85\textwidth}
        \centering
        \includegraphics[width=0.9\textwidth]{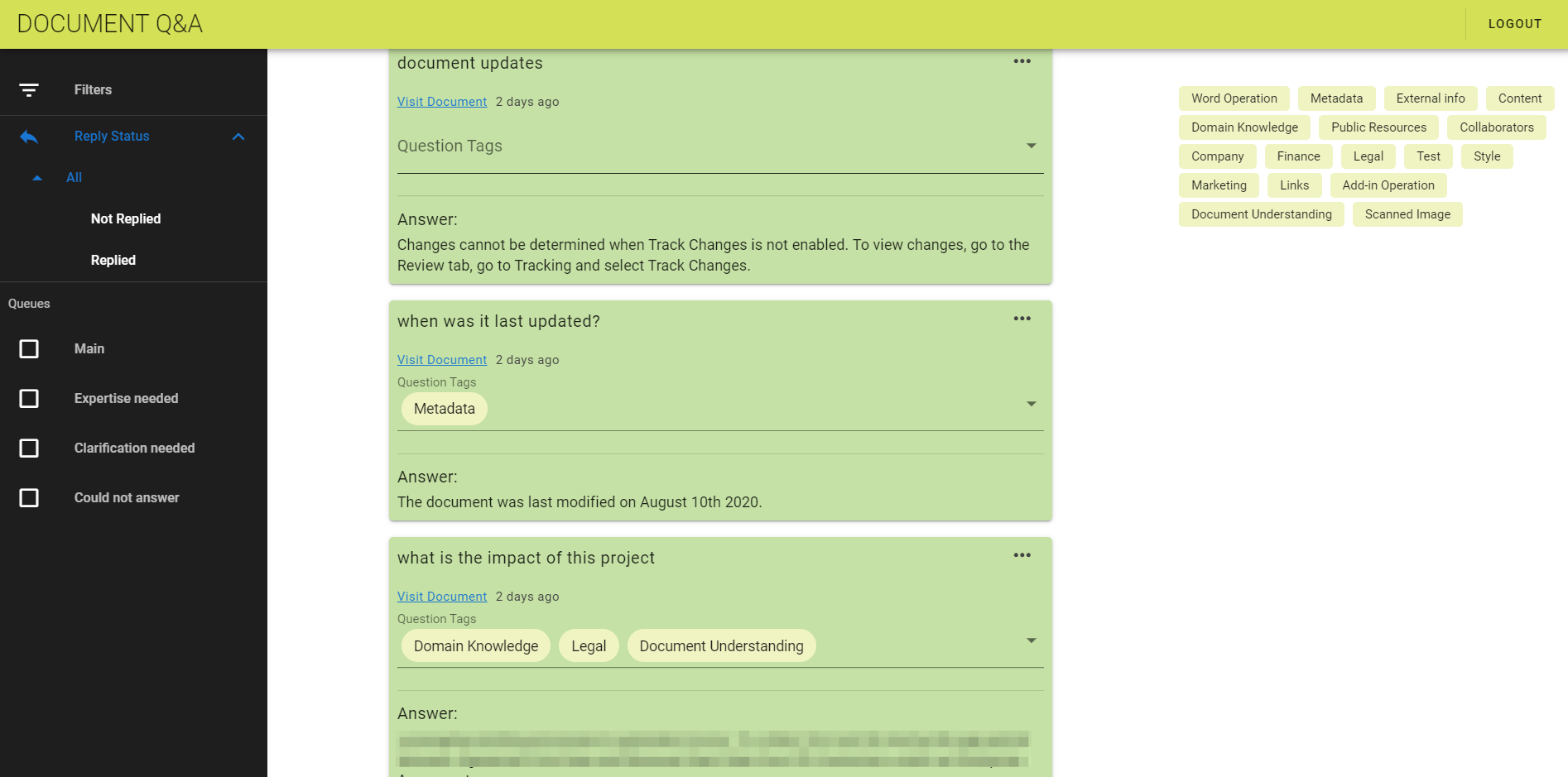} %
    \end{minipage}\hfill
    \caption{The document Q\&A worker view. Each question that was submitted via the Q\&A addin would be visible in this system and workers could visit the document to answer the question. Additionally, workers could move the questions to different queues (filters seen on the left) or assign tags to each question (seen on the right).}
	\label{fig:phase_two_worker_view}
\end{figure}

The knowledge workers' view contained all the questions submitted by users (see Figure~\ref{fig:phase_two_worker_view}).
As each question arrived, the workers performed the first task of determining whether an answer to the question could possibly be provided at all. An example of a question that could not be answered would be if the question appeared to refer to the content or the style of the document but a share link to the document was not provided or that the share link was not valid. The workers would place the question in different queues based on this criterion. Additionally, workers assigned finer-grained tags to questions and iterated over tags already assigned to prior questions as new ones came in. This approach of assigning tags at question arrival helped with the organization of the questions and routing them to the different components of the system. 
In addition, the tags served as a basis for the taxonomy that we developed from the question pool.

The workflow for answering questions was based on the question type.
For questions about the document content, workers would manually access the document using the share link and their personal credentials. They would then copy and paste the question and document content into a custom UI front-end for the ML Q\&A model.
If the AI-provided answer was unsatisfactory upon an initial inspection (e.g., if the selected passage did not seem to answer the question), workers would answer the question by reading through its corresponding document.
Workers answered the questions about metadata either by investigating the metadata captured by the add-in at the time of question submission or by accessing the document through the share link if the requested metadata was not among the contextual information captured by the add-in. Questions about style were similarly answered by accessing the document. Some questions sought external information that was available on public resources. Workers answered these questions by retrieving relevant information using a search engine. If a question requested external information available only within the company, workers used the internal repository of documents shared with employees to find the requested content.

In general, the delay in answers that the AI produced was not much different than those the human workers composed, since the time a question spent in the queue was the dominant source of delay across all categories.

When composing answers, workers imitated the response style of the component to which the question would be directed in a fully automated Q\&A system. For instance, for answering questions that were about content and that the model could not answer, the workers would put together excerpts extracted from the document with minimal change to enhance the flow of the text. For responding to a question whose answer could be found in the internal document repository or using a search engine, the workers would reformulate the query, browse through the document they deemed most relevant, and would copy the excerpt(s) containing the response, similar to how in response to a search query, the answer is automatically extracted from the most relevant document and highlighted to the user. In contrast to most machine generated answers however, if the question was a yes/no question (rather than WH), the yes or no answer was provided first and it was potentially followed by the excerpt on which it was based.
For questions that required reasoning, the workers tried to provide succinct answers to minimize the perceived human interference (e.g., the answer to the question\textit{``how many questions in total?'' (p-2-11)} was \textit{``19''}). Some of the questions in this category however, were formulated in a way that suggested the user expects a response either from a highly sophisticated AI or a human worker. In these cases, the responses also incorporated a superior level of understanding. An example is the following:

Question: \textit{``I pasted in a table from a PPT slide and the table dimensions are larger than the page which makes it cut off. Is there a way to have it auto-scale to the page, without having to manually resize the table?'' (p-2-9)}

Answer: \textit{``Copy and paste the large range of data into Word document, and then select the pasted table, then click Layout > AutoFit > AutoFit Contents / AutoFit Window.''}

For questions that could not be answered, workers would submit a template response that was determined based on the question type and what information was missing. An example of such a template response was: \textit{``Cannot answer this question because the question appears to refer to the [document aspect] and [the problem with access to the document]''}. Depending on the type of the question, sometimes the answer contained additional directions for the user to pursue (e.g., \textit{``Cannot answer this question as it appears to require domain knowledge. The question is likely best directed to the document author, or to collaborators.''}). These template responses were developed as questions came in.

The median response time in this phase was 17 minutes, with an average of 1.5 hours. Even outside work hours, we tried to keep answering questions, but those questions sometimes experienced longer delays. 

At the conclusion of the two week period of the study, 41 users had submitted a total of 130 questions asked about 61 distinct documents. We then distributed an end-of-study survey to participants asking about their experience with using the system. The survey questions asked for examples of good and poor answers that they received from the system. For each example, the survey asked about the type of question they had asked and the delay in receiving the response. The survey also asked about the overall satisfaction with the system (5 point Likert), how the system could be improved, and if they would recommend using the system to colleagues (5 point Likert). A total of 31 participants from this phase completed the questionnaire.


\subsection{Taxonomy Development}
To develop a taxonomy of the types of questions with which people need support when working on business documents, we combined the questions we collected from both phases of the study. The questions from Phase 2 already had preliminary labels assigned to them as the workers had organized the questions to decide the course of action for answering them. We divided those submissions from Phase 1 that contained more than one question into idea units. With this division, the total number of questions from Phase 1 and Phase 2 was 272. A member of the research team then used open-coding to inductively develop codes that encompassed thematically related idea units.
The categories assigned to each question were not a property of only the question but also of the document about which the question had been submitted. For instance, the question \textit{``What is client sdk?''} can be a content-related question or one seeking external information depending on whether ``client sdk'' is defined in the document or simply mentioned in the document in some context.

Through subsequent passes, the labels with too much overlap were consolidated and the ones showing distinct ideas were further split into separate categories. Another member of the research team was then trained on the categories and used them to label a randomly sampled set of 80 idea units (29\% of the total). Cohen's Kappa for the high-level categories was 0.86. To determine the inter-rater reliability for the subcategories, we assigned each category that did not have nested subcategories a subcategory with a value equal to the value of the category. Cohen's Kappa for the subcategories was 0.75. Both Kappa values exceeded the recommended threshold for accepting the results~\cite{landis1977measurement}.

\subsection{Participants}

We recruited participants in both phases by randomly sampling email addresses from, and sending invitations to, employees of Microsoft. A total of 59 and 41 users participated in the Phase 1 and Phase 2 studies respectively. Across both studies, 37\% of participants were female. The medians of reported age and highest education achieved were 35-44 and Bachelor’s degree. Participants came from a diverse set of roles at the company including software and hardware engineering, sales, content writing, program management, finance, communications, etc. The medians for how frequently the participants read or edited Word documents were both ``a few times per week''. Participants copy-edited Word documents less often (a few times per month).  

We compensated participants from the Phase 1 study with a base amount of \$20 in the form of an e-gift card. To encourage more involved participation, for each question that a participant submitted in each day, they would secure an entry into a raffle for 5 gift cards, each with a value of \$50. We limited the number of entries per day for each participant to 25.
The compensation for Phase 2 of the study was \$20 e-gift cards. Although the duration of Phase 2 was longer than Phase 1, receiving answers to questions about documents was another form of value that participants received from the study.

\section{Results}
\subsection{Taxonomy of Question Types}
We categorized the questions of our study based on what type of information they sought and where that information was available. Table~\ref{tab:question_types} shows the full taxonomy as well as examples for each category. Throughout the paper, where we present participants' questions, we identify them with a string of the form `p-' + phase number + participant number.

Participants used different languages for articulating their information needs. Some queries were rather verbose---\textit{``Is there a way to resolve comments and then make them not show up in the comments pane? Sometimes I end up just deleting the comment b/c I want to clean it up before I send it for final review from a key, leadership stakeholder.'' (p-2-9)}, and others, short---\textit{``app service blog'' (p-2-26)}. Both types of queries can cause challenges for AI systems to interpret user intents~\cite{kumaran2009reducing}.


Among the questions that we categorized as \textit{Workflow \& Operation Help}, some were posed in the format of help queries and others were commands or direct requests from the assistant. The help queries were either about general procedures---\textit{``How to create a table?'' (p-1-39)} or specific to the user's context---\textit{``Why is there the warning "Upload blocked" as I already have a copy in my local storage?'' (p-1-11)}.
The command type submissions were more common in the Phase 1 study in which participants were free to imagine an assistant that could help with any of their document-related tasks. In the Phase 2 study however, because the system afforded help with document consumption and not manipulation, participants rarely submitted commands. The majority of the commands that the users submitted requested features that were not implemented in Word---\textit{``Find all the pink words and review for modifications'' (p-1-8)}.

A number of the questions that we categorized as seeking \textit{External Information} were tied to the content of the document--- \textit{``Love for the tool to find market research data regarding the topic. Both in our own data and from known trusted research companies online.'' (p-1-52)}. The reason for this categorization however, was that the information that they sought resided outside the document.

Some of the categories that emerged from the questions submitted by our participants have parallels in the taxonomy developed by Ter Hoeve et al~\cite{ter2020conversations}. For example, the questions that they classify as factoids appear to be questions seeking external information but only on public resources. The questions that they categorize as mechanical, copy-editing, or navigational were commands or requests related to the operation of Word in our study. The rest of document-related questions that they obtained we categorize as questions related to the content. Their taxonomy however, did not contain questions of other types present in our study, for instance those related to style or seeking external information from collaborators. One reason for this could be because question generators in their study did not in fact work on the documents but rather assumed familiarity with them. Another could be that those documents were in final published form and not in development. Yet another reason could be the context of their study which emphasized AI-based assistance while we explained to our users that the AI would be complemented by human intelligence.

\subsection{AI Success Rate}
Although the AI model was invoked for every content related question in Phase 2, success was limited to three factual questions, representing 25\% of the factual questions, 7.3\% of the content questions, and 2.3\% of all questions in this phase. Of these answers, two were relayed just as the model produced them and one was modified by the human workers to contain more information. In the other cases, it did not provide an answer at all or the answer was considered not relevant by our workers.

\subsection{Difference in the Distribution of Questions by User Role}
\label{section:user-roles}
To gain insight into whether the distribution of questions differs by the user's role in the document, we investigated responses and contextual information collected by the add-ins. 
In the Experience Sampler questionnaire of Phase 1, we directly asked about the user's primary role in the document. We also collected metadata about the users' activities. Specifically, if the user was the creator, we assigned them the role \textit{author}. If the user had not created the document but had edited or commented on it, we assigned them the \textit{reviewer} role. If the user had done neither, we labeled them a \textit{reader}.


\begin{small}
\begin{longtable}[c]{|p{1.3cm}|p{3.3cm}|p{3.8cm}|p{4.1cm}|}
\caption{Taxonomy of the types of questions with which users need support when working with business documents.}
\label{tab:question_types}
\\
 \hline \rowcolor{gray!28} \multicolumn{1}{|l|}{\textbf{Category}} & \multicolumn{1}{l|}{\textbf{Definition}} & \multicolumn{1}{l|}{\textbf{Subcategories}} & \multicolumn{1}{l|}{\textbf{Examples}} \\ \hline 
 \endfirsthead
 
 \multicolumn{4}{c}%
 {{\tablename\ \thetable{} -- continued from previous page}} \\
 \hline \rowcolor{gray!28} \multicolumn{1}{|l|}{\textbf{Category}} & \multicolumn{1}{l|}{\textbf{Definition}} & \multicolumn{1}{l|}{\textbf{Subcategories}} & \multicolumn{1}{l|}{\textbf{Examples}}\\ \hline 
 \endhead
 
 \hline \multicolumn{4}{|r|}{{Continued on next page}} \\ \hline
 \endfoot
 
 \hline \hline
 \endlastfoot

 \textbf{Metadata} (N=32) & Questions about one or more of the following: actions (e.g., edits or comments), time of actions, actors, and document's properties (e.g., word count) & --- & \textit{``Was this file's location moved in the last 6 months?'' (p-2-6)}\newline
 \textit{``When was the last time this info was updated?'' (p-2-25)}
 \\
 \cline{1-4}
 \textbf{Workflow \& Operation Help} (N=61) & \multirow{2}{=}{ Questions about how to or commands to perform a specific task in Microsoft Word or the Q\&A add-in} & \textbf{Help Query} (N=54) & \textit{``How can I tell if an embedded Excel table is current and could a reviewer view the data source?'' (p-1-42)} \\
 \cline{3-4}
 \ & \ & \textbf{Command or Request} (N=7) & \textit{``Show me my unresolved comments.'' (p-1-47)}
 \\
  \cline{1-4}
 \textbf{Content} (N=56) & \multirow{2}{=}{Questions that could be answered from the content of the document} & \textbf{Factual}: A question the answer to which can be retrieved as a passage from the document (N=19) & \textit{``How much morale money did everyone get in June?'' (p-2-32)} \\
 \cline{3-4}
 \ & \ & \textbf{Reasoning}: A question answering which requires complex reasoning and/or use of external information (N=25) & \textit{``Are there any action items in this document?'' (p-2-14)}\newline
 \textit{``How many different databases are mentioned''? (p-1-4)}
 \\
\cline{3-4}
 \ & \ & \textbf{Overview}: Refers to the document as a whole---including document type, topic, impact of the document, etc. (N=9) & \textit{``What is the focus of this document?'' (p-2-35)}
 \\
 \cline{3-4}
 \ & \ & \textbf{Summary}: A special case of overview questions, seeks the summary of the document or a section of the document (N=3) & \textit{``Can you summarize the main points of each section of this document?'' (p-1-13)}
 \\
 \cline{1-4}
 \textbf{Written Style} (N=10) & Questions about the organization, syntax, or semantics of the text or the language of the document & --- & \textit{``Are there any statements in this document that could be unclear to the reader, or are not inclusive in nature?'' (p-1-8)}\newline 
 \textit{``Does this document contain unnecessarily wordy paragraphs?'' (p-1-50)}\\
 \cline{1-4}
 \textbf{Visual Style} (N=9) & Questions about the document's current formatting, layout, typography, etc. & --- & \textit{``Why are there weird spaces in my document?'' (p-1-1)}
 \\
 \cline{1-4}
 \textbf{External Information} (N=72) & \multirow{2}{=}{Questions seeking information that is external to the document. This information could be available:} & \textbf{On public resources}, e.g., public documentations (N=21) & \textit{``What is executive bias?'' (p-2-32)} [Asked in the context of a document that mentioned executive bias but did not define it.]\\
 \cline{3-4}
 \ & \ & \textbf{Within the company}, e.g., usage data of specific apps (N=12) & \textit{``Where can I find a template of a communication planning document?'' (p-1-42)}
 \\
 \cline{3-4}
 \ & \ & \textbf{To the author or collaborators of the document} (N=22) & \textit{``What is our target date to publish this document?'' (p-1-37)}
 \\
 \cline{3-4}
 \ & \ & \textbf{To experts} (N=15) & \textit{``What are some general topics that are covered when building a training agenda for a specific program?'' (p-1-54)}
 \\
 \cline{3-4}
 \ & \ & \textbf{Across applications} (N=1) & \textit{``Can word check a date and time against my outlook calendar for conflicts?'' (p-1-18)}
 \\
 \cline{3-4}
 \ & \ & \textbf{Other} (N=1) & \textit{``How old is my son?'' (p-2-26)}
 \\
  \cline{1-4}
 \textbf{No Question} (N=28) & Users specifying they did not have a question at the time of submitting the Experience Sampler or the Q\&A form. These submissions were more common in the Phase 1 study where we asked users about their questions at random points in time. & --- & \textit{``I do not have any questions at this time.'' (p-1-10)}
 \\
  \cline{1-4}
 \textbf{Unknown} (N=4) & Questions that may be either content-related or seek external info but that we did not have access to the content of the document to tease them apart & --- & \textit{``What activities are needed?'' (p-2-34)}
 \\
 \hline
\end{longtable}
\end{small}

In most cases in Phase 1, there was good (71\%) agreement between self-reported and metadata-assigned roles. In cases of disagreement, we found that some users who had edited the document had labeled themselves as readers, while others labeled themselves as reviewers -- likely signaling their perceived degree of contribution.   

Given the high degree of agreement in Phase 1, we opted to drop the role question from Phase 2 of the study, instead relying exclusively on metadata. This kept the UI streamlined around asking and answering questions efficiently, yielding an experience resembling what a user would expect to see if such a system were deployed in the wild.

Analyzing data from Phase 1 and Phase 2 together, we  performed a Chi-squared test of independence on the contingency table of question categories and users' metadata-assigned roles. We excluded queries of the type \textit{No Question} or \textit{Unknown} from the data. The test revealed that the distribution of questions is in fact not independent of the user's role ($\chi^2(10)=37.78$, $p<0.001$).

\begin{figure}[!t]
    \centering
    \begin{minipage}{0.9\textwidth}
        \centering
        \includegraphics[width=0.9\textwidth]{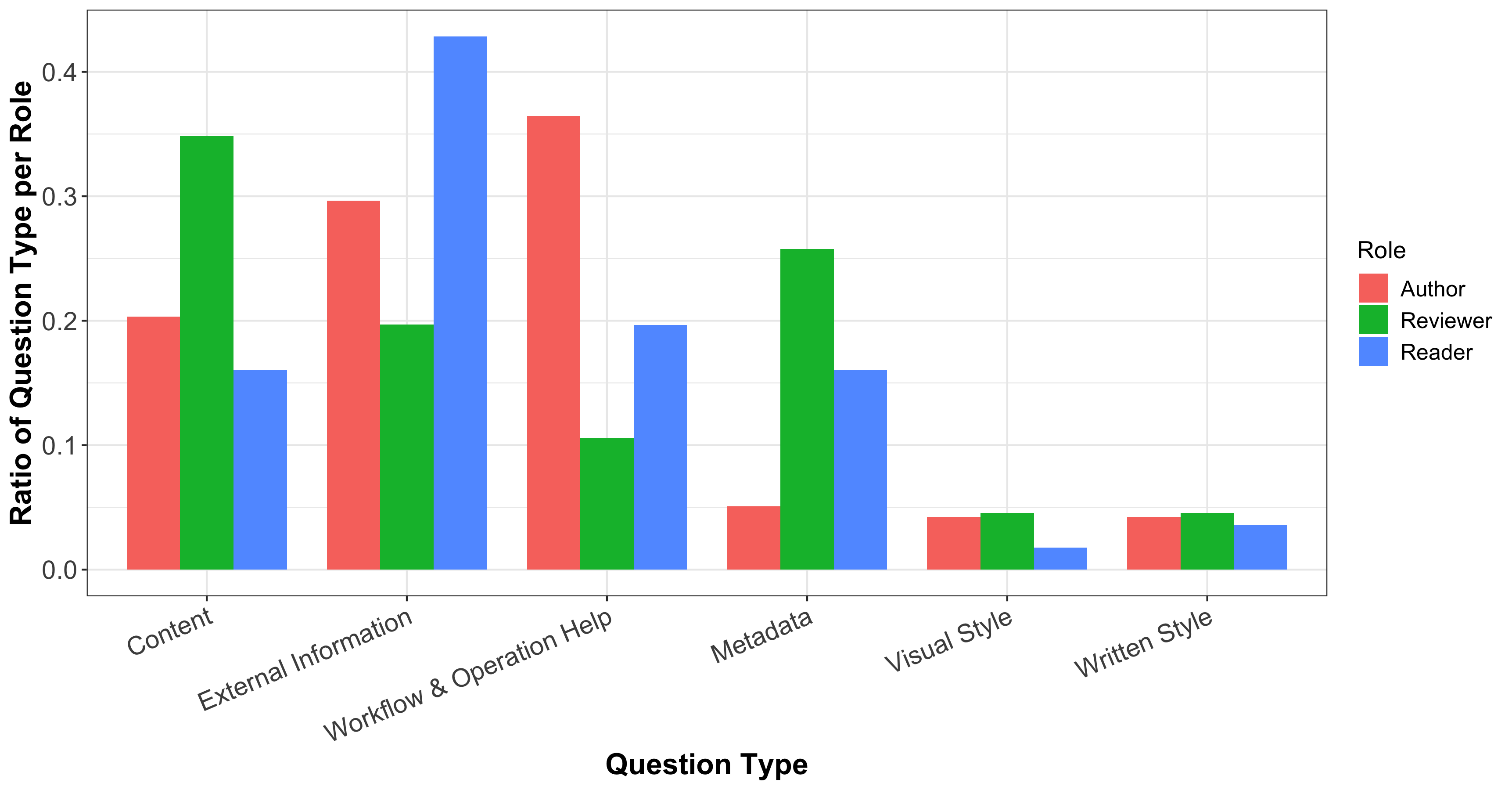} %
    \end{minipage}\hfill
    \caption{Distributions of question types by the role of the user in the document. Each bar shows the ratio of the question type relative to all questions asked by users with similar role.}
	\label{fig:question_types_by_role}
\end{figure}

Figure~\ref{fig:question_types_by_role} displays how the distributions of questions across categories vary by the user's role. Because the numbers of questions asked by authors, reviewers, and readers of documents are different ($N_\textit{Author}=118$, $N_\textit{Reviewer}=66$, $N_\textit{Reader}=56$), the bar for each question category and each role is normalized by the number of all questions asked by users with a similar role. The figure suggests that the questions that authors ask are more often concerned with performing specific operations in Word---\textit{``How to save as pdf without showing the comments?'' (p-1-11)} or finding information external to the document---\textit{``How would one approach a point of view paper?'' (p-1-7)}, both of which help with authoring tasks. Interestingly, readers also ask more questions about external information rather than the content of the document, although it is potentially the content of the document that gives rise to such questions---\textit{``Who's the competition?'' (p-1-41)}. Reviewers however, are more concerned with the document's content---\textit{``...What action [is] needed from me?'' (p-1-35)} or its metadata---\textit{``Where is the last change?'' (p-2-10)}, with both types of questions helping the reviewer find the information that is relevant to them. 
As expected, readers are not concerned with the visual style of the document as often as authors or reviewers, although the datapoints in this question category may not be enough for generalization.

\subsection{Skillsets and Roles of Human Respondents}

Many of the submitted queries required a level of understanding, analysis, or need of a knowledge base that would be challenging for even state-of-the-art ML systems. To answer these queries therefore, a document Q\&A assistant would need to employ human intelligence. In our study, the type of human respondents that were needed to answer participants' questions included the author or document collaborators---\textit{``When is this slated for release?'' (p-2-16)}, experts in various domains---\textit{``Does [cloud service product] use my [OAuth provider] account?'' (p-1-6)}, and human workers with a more general skill set---\textit{``Can the assistant make recommendations to other ways to phrase my questions?'' (p-1-12)}.

The types of questions that needed human intelligence were not limited to those that asked for external information or an opinion, but also included those that referred to the content of a document but required domain expertise to answer, for instance, in the domain of finance, legal, marketing, or software engineering.
The desire for connecting with colleagues familiar with the document or domain experts within the company explicitly surfaced in Phase 1, where participants had also indicated who could answer their question. We present a sampling of questions, and suggested human respondents below:

\textit{``What information is missing? Do the readers understand the content and what questions do they have?''---suggested respondent: \textit{``People who have written similar documents or documents on this topic''} (p-1-27)}

\textit{``Why aren't images in my clipboard being pasted/transferred to the word document?''---suggested respondent: \textit{``Someone from the [TEAM NAME] team that can troubleshoot this.''} (p-1-46)}


The end-of-study survey responses also indicated the need for domain experts in such Q\&A systems where the most cited question type for which participants stated they had received an unhelpful answer was one needing domain expertise. When the knowledge workers of our study received a question answering which required expertise they did not possess, they marked the question as unanswerable, explaining the reason.

\subsection{Users' Experience with the Q\&A System}

Overall, users were fairly satisfied with the answers they received from the Q\&A system ($\mu=3.45$, $\sigma = 0.98$). Only 4 out of the 31 end-of-study survey respondents rated the answers unsatisfactory (rating of 1 or 2). We explored participants' free-text responses to understand what they liked or disliked about the system. Participants  saw value in the tool helping them be more efficient at their task ---\textit{``It helped me by providing answers quickly, I could have probably checked myself some of them by going through previous versions of the documents etc., but this was more efficient.'' (p-2-29)} or find solutions or workflows appropriate for their particular context (N=3)---\textit{``I was able to ask questions about a few different capabilities that I was unfamiliar with before, which was nice.'' (p-2-9)}. A number of participants (N=4) stated that they would have liked to know about possible use cases of the tool---\textit{``I think it would be helpful if the system could present some sample questions so the user could know what type of answers the system could provide.'' (p-2-14)}. We had deliberately withheld examples of usage from users because we did not want to prime users to ask questions of a particular type. Some participants asked for a faster response time---\textit{``More expedient answers'' (p-2-5)}. Others (N=2) had concerns about privacy---\textit{``be able to search some properties of a document without full access'' (p-2-38)}.

We performed an exploratory analysis to understand if the delay in responses had affected user’s satisfaction with the tool. We developed three linear models with satisfaction rating as the dependent variable and a measurement of response times a participant had experienced (mean, minimum, and maximum) as the independent variable. We did not observe a significant effect of time on satisfaction ratings ($\beta=0.000, p=0.64$ for mean; $\beta=0.000, p=0.57$ for min; $\beta=0.000, p=0.83$ for maximum). When examining participants’ responses, we found that, interestingly, what is perceived as a long delay differs across participants. In response to how quickly the answer to their question arrived, one participant said \textit{``Quickly, I think within a couple hours.''}, while another wished for a faster response time---\textit{``There was a delay in the email sent to me. Took about 5 minutes''}. This variance in how long a delay is acceptable may be due to the difference in question types and their complexity. We examine the relationship between question types and their expected response times in the next section. 

\subsection{How Response Time Varies by Question Type}
\label{section:additional-analyses}

\begin{figure}[!t]
    \centering
    \begin{minipage}{\textwidth}
        \centering
        \includegraphics[width=\textwidth]{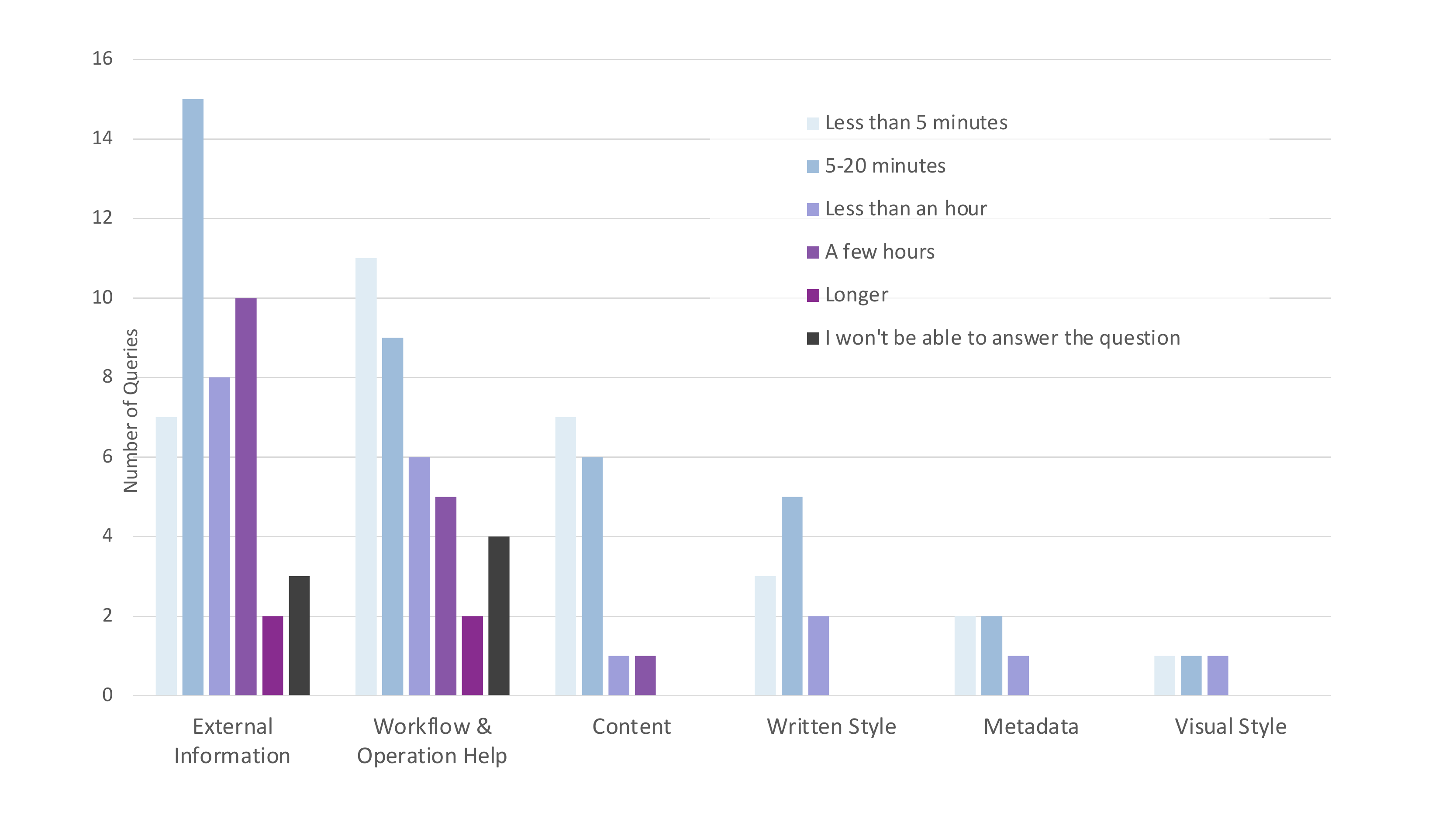} %
    \end{minipage}\hfill
    \caption{Distributions of self-reported response time by the type of questions. Each bar shows the number of the questions in the specified category that could be answered within the indicated response time.}
	\label{fig:question_types_by_response_time}
\end{figure}

To further examine what types of questions a document Q\&A system could be most valuable for, we explored how the type of questions affects the time it takes users to answer them. To do so, we probed the data from the Experience Sampling phase in which participants submitted their estimate of the time it would take them to answer their query along with the query itself. We excluded the queries of the type \textit{No Question} or \textit{Unknown}, leaving 115 queries for analysis. Figure~\ref{fig:question_types_by_response_time} shows the distribution of self-reported response times across question categories. 
The figure shows that most content-related questions can be answered by the participant in a relatively short time (less than 20 minutes). Interestingly, the questions that participants indicated would take them hours to answer or that they could not answer at all were related to External Information or Workflow \& Operation Help.

\section{Discussion}

The results of our studies contribute empirical understanding of users' information needs when working with their documents. Understanding the types of assistance that users require is a stepping stone for designing systems that can provide that assistance. 


The human-in-the-loop document Q\&A prototype in our study was a first attempt at the kind of document digital 
assistance
that we envision. It served as a technology probe to understand the needs of users in a real-world setting, examine the feasibility of generating answers with the aid of AI and other humans, and understand the design requirements of such a technology~\cite{hutchinson2003technology}.
Using a document Q\&A prototype and collecting questions in-situ as users were working on their documents naturally and when they had questions, enhanced the ecological validity of our study with respect to both the nature of questions asked and the frequency with which such a tool would be used. While this data collection method did not result in a large number of questions per participant, it yielded a sample that represented actual questions in the wild. Indeed, before the onset of the study, some participants told us that they anticipated accessing only a few Word documents during the study period.

In the context of our study, because human operators had access to the entirety of the documents through the share links that users provided, in asking questions, users were not bound to the capabilities of current digital assistants. This versatility allowed users to receive help with the documents they worked on and us to collect a variety of questions before attempting to develop a document Q\&A system which may not align with users' actual needs.

\subsection{Question Types}
One surprising finding of our study was that despite the significant investment in research on extractive question answering (used for factual questions)~\cite{rajpurkar2016squad, nguyen2016ms}, many questions are in fact not limited to the content of the document. Moreover, the majority of the content-related questions that the participants asked could not be answered by returning an excerpt from the document. The shortcomings of existing question answering models could be partly attributed to the datasets they use for training which consist of questions and answers generated by crowd workers or search queries on documents on the web. These queries have been on finalized public documents authored by someone other than the query generator and on which query generators have had little prior knowledge. Therefore, most queries obtained in this corpus lack the diversity of attributes that can influence the type of content-related questions. 

One such attribute is the stage or state of the document at the time the question is asked. For instance, questions about style or metadata are not often asked on an already published document.

Another such attribute is the role of the user asking the question, which can potentially affect the type of content-related questions they ask, similar to how user's role affects their overall types of questions, discussed in Section~\ref{section:user-roles}. For instance, in the cases where the user is an author of the document, they already have enough familiarity with the document and the context of the information they seek to look for it using navigation or the ``find'' tool instead of submitting their query to our system which would take longer to respond. As one participant said:
\textit{
``[M]ost of the questions I had could either be found directly in the doc with a quick Ctrl+F or, for metadata, just looking in file Properties dialog box (right click on file)...'' (p-2-25)}. In the cases where the user is reading a document they have not authored, their content-related questions may more often involve an overview or finding an answer to a specific question that requires some degree of reasoning. In cases where factual questions were asked, the end-of-study survey responses suggest that the participant may have been posing questions to familiarize themselves with the system:
\textit{``Even before asking questions that'd be actually useful, I began by asking questions that test its ability.'' (p-2-37)}. 

Our finding that even state-of-the-art Q\&A models fall short of their promise to answer factual questions in the domain of business documents does not suggest that a digital assistant will be unhelpful and that we should abandon efforts to build such an assistant in this domain. On the contrary, it points to directions that researchers and designers of these tools can pursue to overcome the challenges of this domain. One is that such tools, at least for the present, need to employ a combination of automation and human intelligence. The questions, documents, and human generated responses of a hybrid Q\&A system can serve as ecologically valid training data for building future Q\&A models. With our results showing that the effectiveness of such models differs drastically across hypothetical data and data collected in-situ, we call on future work to pay special attention to the process of collecting training data.

\subsection{Opportunities and Challenges of Incorporating Humans in the Loop}
Another insight from the range of the questions we received was that a number of them required human assistance from the document's authors, domains experts, or workers with more general skill sets. A document assistant would therefore need to classify and route questions to the appropriate respondents. This component of the system would be akin to IM-an-Expert, a social Q\&A system which located and contacted potential respondents with expertise or interest in a subject matter \cite{richardson2011supporting, white2011effects}. While questions submitted to a human-in-the-loop document Q\&A tool may experience delays, the delay could be justified if the tool enables users to be more efficient or help them accomplish tasks that they could not otherwise perform. Indeed, in Section~\ref{section:additional-analyses}, we observed that there are certain types of questions such as those seeking information external to the document or workflow and operation help which would take participants hours to answer or that they would not be able to answer at all. These questions are of the type that could be routed to and handled by other human respondents.

In addition, the delays in responses returned from our Q\&A system were most often due to human availability as opposed to the time it took to complete the task once a human was engaged. In a system where a human is always available we may see a significant decrease in response time. Furthermore, the long-term vision for nearly all the question types is to be accommodated using automated assistance, enabled by collecting in-situ training data from human-in-the-loop Q\&A systems, which will most likely yield even faster response times.

Many questions were beyond the capabilities of the document Q\&A model used in this study. Although human respondents are necessary for some of these questions, others have great potential for automation, if only the model could consult other resources. For instance, for answering the questions seeking external information available on public resources, and for some questions about the operation of Word, a Q\&A system could make use of the research in information retrieval and search. Questions seeking information within the company's documents could use the same approach while also leveraging enterprise knowledge bases, together with rich contextual information such as the role or expertise of candidate documents' authors, their organizational distance from the user asking the question, or the authors' ownership of similar documents in addition to the content of the documents. Finally, questions on metadata could be answered by mapping the questions to the correct application interface (API) calls to retrieve the necessary information \cite{su2017building}.
We have summarized several pathways for automating the handling of different question types in the business document domain in Table~\ref{tab:automating_QA}.

While many document-centric requests can be automated and therefore accomplished without the user having to share their content with someone who does not otherwise have access to it, there still exist information needs that cannot be satisfied by pure automation. Because the knowledge workers in this study were the researchers on the project and employees of the same company as the users, most users agreed to sharing the content of their document with the workers. However, privacy may be a challenge in cases where users work with confidential documents or if workers are recruited from outside the company. This issue was addressed in a similar hybrid system, Calendar.help \cite{cranshaw2017calendar}, by having human workers signing a non-disclosure agreement and by designing microtasks that include only the information needed to complete the scheduling task. Future work can investigate other approaches to maintaining confidentiality of documents and the user's privacy for instance, by chunking the document into different segments and assigning each to a different knowledge worker in a manner similar to the context-free microtasks explored in \cite{iqbal2018multitasking}. In fact, research has examined how to withhold personally identifiable information in images from crowdworkers by showing only small segments and iteratively zooming out to identify visual information and by leveraging workers' prediction of adjacent segments that are not displayed \cite{lasecki2015preserving, kaur2017crowdmask}. In the context of documents, some tasks such as stating the impact of a document or critiquing its overall language style may require a holistic view. However, it is conceivable that teams of knowledge workers could dynamically be organized similar to flash teams in \cite{valentine2017flash, retelny2014expert}, where outputs or summaries provided by some teams are given as inputs to others, eliminating the need for any one worker to have access to the specifics of the document.

\begin{footnotesize}

\begin{table*}[!htb]
\caption{Opportunities for automating the handling of different types of questions in the business document domain.}
  
\label{tab:automating_QA}
\Description{Fully described in the text.}
\centering
\begin{tabular}{|p{2.7cm}|p{10.3cm}|}
  \toprule
  \rowcolor{gray!16} Question Type & Opportunities for Automation  \\
  \midrule
   Metadata & Building natural language interfaces to  APIs~\cite{su2017building} that handle metadata\\
  \cline{1-2}
  Workflow \& Operation Help & Re-purposing search engines or information retrieval approaches for this domain \cite{bota2018characterizing}, routing questions to human respondents skilled in Microsoft Word to answer help queries (similar to social Q\&A tools such as ~\cite{chilana2012lemonaid} for web applications), using an embedded tool powered by the crowd to accommodate commands (similar to the Human Macro feature in Soylent~\cite{bernstein2010soylent})\\
  \cline{1-2}
    Content & Using classifiers to triage the type of external information sought, leveraging research in information retrieval and extractive Q\&A for handling factual questions and document summarization for summary type questions~\cite{ter2020conversations, yao2017recent}, using a social Q\&A system to locate and contact human respondents with appropriate skillsets for answering overview or reasoning type questions\\
   \cline{1-2}
    Written Style & Using a social Q\&A system to contact human respondents with appropriate or specific skillsets, using an embedded tool powered by the crowd for accommodating proofreading requests (similar to the Proofread feature in Soylent~\cite{bernstein2010soylent}), retrieving answers to general syntactic questions in knowledge bases such as \textsf{english.stackexchange.com} \\
   \cline{1-2}
   Visual Style  & Building natural language interfaces similar to those targeted for GUIs~\cite{todi2021conversations}, using a social Q\&A system to contact human respondents with appropriate skillsets\\
   \cline{1-2}
    External Information & Using classifiers to triage the type of external information sought to determine the corpus of search, leveraging information retrieval techniques within the corpus, routing questions to the appropriate human respondents similar to IM-an-Expert~\cite{richardson2011supporting, white2011effects}, enhancing the Q\&A social system using enterprise knowledge bases enriched with contextual information such as employee roles, expertise, inquirer's and potential respondent's organizational distance from each other, and history of potential respondents working on similar documents as one on which a question is asked\\
  \bottomrule
\end{tabular}
\end{table*}
\end{footnotesize}

\subsection{Opportunities for Human-in-the-loop Document Manipulation}
Our study was a technology probe to understand how current document Q\&A models can be leveraged to help with consumption of business documents, how well these models deliver their promise, and to gain insight into opportunities of employing human intelligence to fill in the gaps of current technology. Therefore, the prompts that we presented to the participants in both phases  nudged them to ask questions that would help them with consuming the document or accomplishing a particular task, rather than delegating a task to the assistant. While we received some questions in the form of commands directed to the assistant in the Phase 1 study, the majority of questions were not task-oriented commands as studying document manipulation in this domain was outside the scope of our work and would require a study design targeted for this context. One difference between the context of accommodating queries in consumption vs manipulation is that in consumption, the question and the answer can be private to the user, whereas when manipulating a document, the change will be visible to all the users who have access to the document. One question that arises in this context is whether the changes should appear to have been made by the user who made the request or by the AI. This design decision would likely impact the types of tasks with which users will ask for help. The types of requested tasks will also likely differ by whether the document is shared with others.

A challenge of accommodating manipulation in the domain of business documents is that the tasks, at least for the present, will likely need to be completed solely by human workers, as the scope of automation in this domain is very limited. Prior work has already investigated crowdsourced tools to help with modifying general documents~\cite{bernstein2010soylent}. Future work should conduct a need-finding study to investigate whether in the business domain, users are willing to outsource modifications to their documents to others who may not have the required background knowledge or domain expertise, or may simply not know the context of the project. A need-finding study should also examine whether using such a system can impact the collaborators’ perception of the quality of the completed task or the user who requested the task. We have added these points to the Discussion as areas that future work can investigate.

\section{Limitations and Future Work}
In this work we focused only on Word documents as a common document type where users can author, copy-edit, or read content. 
Indeed, the organization where we deployed our studies primarily uses Word for business documents. The focus on Word rather than e.g., PDFs or web pages also allowed us to obtain questions about documents at various stages of development, not just finalized manuscripts.
Future work can investigate the types of questions that arise when users interact with other file types such as PDF, Excel, and PowerPoint. Because each file type is used for different purposes (e.g., Excel documents for long-term book-keeping \cite{jahanbakhsh2020effects}) and possibly containing content at different levels of abstraction, the extent to which question answering in these documents can be automated and the kinds of expertise knowledge workers need may be different from the Word documents in our study.

Because all questions had to first pass through the knowledge workers' system, the majority of the responses that participants received from the Q\&A system had a delay. Therefore, it is possible that participants would have asked different questions if the responses had been provided instantaneously.
Another property that could have conceivably impacted the types of questions users asked is the quality of the answers that they received. To understand if the types of questions by a user changed over time as they gained familiarity with the system, we examined the questions that were posted by the same user both across different documents or on the same document. We observed that some users posted multiple questions in succession and close together in time, e.g., in the span of a few minutes. Although these users would realize that the system does not provide answers instantaneously, their cluster of initial questions would not be impacted by the expectation of a long delay, their perceived capabilities of the system, or the quality of the answers.
In fact, although we had specified in the consent form that there may be a delay in the responses that participants would receive from the Q\&A system, some end-of-study survey responses indicated that a number of participants had in fact not noticed this point and had asked their first few questions expecting instantaneous answers---\textit{``I was initially confused by the delay of asking the question in the document and then waiting for an email that told me to go back to the document. It seemed a bit redundant to get an email about it vs. just a notification in the Word doc itself and telling me it was working on it or something. For a plug-in, however, I would expect less of a delay.'' (p-2-18)}. Therefore, the first few questions from these participants could help with the generalizability of our results.

The delay or the quality of answers however, may have influenced those users who submitted questions after receiving answers from the system. Upon examination, we found that in many instances, when users received answers to their previous questions, they explored increasingly more sophisticated questions of various types, e.g., content-related, concerning metadata, or seeking external information. For instance, one participant started by asking simple metadata questions (\textit{“who is the author?”}) and proceeded to ask questions on a scanned document that needed not only complex reasoning, but also optical character recognition (OCR): \textit{“what is the total score?”} on a document where handwritten scores were given to each question.
This finding suggests that the answers may have in fact encouraged users to be liberal with the types of questions they wished to subsequently ask. This exploration could be due to users gaining confidence that the system can in fact handle the types of questions with which they need support or could be because they wished to test the limits of its abilities.

Nevertheless, the characterization that we present in the paper also includes the questions users submitted in the Experience Sampling phase, where participants could imagine a sophisticated system
with any or no delay. The set of questions in that phase of the study could further help with generalizability of our results, e.g., to settings where not all questions necessarily experience a delay.


It is conceivable that the types of questions about a document may vary with the document type. In Phase 2 where we had access to the documents, we observed that the document distribution was in fact very varied and included project proposals and timelines, value propositions, design specifications, service instructions, management training, protocols, FAQs, whitepaper reports, strategy planning, customer feedback, research findings, etc. from various domains. With this diverse set of document types, investigating the relationship between types of questions and types of documents would require collecting far more questions by running the study for a long time.

The setup of our Q\&A system was such that users submitted one-shot questions as there were no affordances for following up on previously asked questions. Future work should investigate whether the types of questions that users ask a document Q\&A system differ if the system provides the users with the means for following up on their previous questions or multi-round conversations with the assistant.

Another area for future work would be to explore the context of user needs including when users need different types of assistance with their documents, what they do before the seek help, and what they do after they receive answers. To minimize concerns about the confidentiality of the business data, the contextual information that we collected in our study concerned the metadata of the document (e.g., file size, last modified date, etc.) and not the content. For the same reason, we did not track a user’s modification of the document before or after the user posted a question; the metadata was collected upon the user's submission of a question. Therefore, given the data we collected, we cannot examine the context of user needs. Future work can study this question through interviews and user-produced logs.





\section{Conclusion}
We studied users' information needs when working with their business documents as a first step towards building document assistants that can handle a variety of user requests. To understand users' actual needs, it was important to collect their document-centric questions in-situ. Therefore, we conducted two user studies. In the first study, we performed experience sampling of users' questions via a Microsoft Word add-in as users were working with their documents. In the second, users submitted their questions via an add-in and received answers from a human-in-the-loop document Q\&A system that complemented a question-answering AI with human intelligence. We characterized the distributions of questions and observed that the types of questions do indeed vary by whether the user is an author, a reviewer, or a reader of the document. In addition, the questions gave us insight into what types of request can be automated and whether particular skillsets or roles within the document are needed from human respondents
in a document digital assistant that is co-powered by artificial and human intelligence.



\bibliographystyle{ACM-Reference-Format}
\bibliography{bibliography}


\begin{thebibliography}{65}


\ifx \showCODEN    \undefined \def \showCODEN     #1{\unskip}     \fi
\ifx \showDOI      \undefined \def \showDOI       #1{#1}\fi
\ifx \showISBNx    \undefined \def \showISBNx     #1{\unskip}     \fi
\ifx \showISBNxiii \undefined \def \showISBNxiii  #1{\unskip}     \fi
\ifx \showISSN     \undefined \def \showISSN      #1{\unskip}     \fi
\ifx \showLCCN     \undefined \def \showLCCN      #1{\unskip}     \fi
\ifx \shownote     \undefined \def \shownote      #1{#1}          \fi
\ifx \showarticletitle \undefined \def \showarticletitle #1{#1}   \fi
\ifx \showURL      \undefined \def \showURL       {\relax}        \fi
\providecommand\bibfield[2]{#2}
\providecommand\bibinfo[2]{#2}
\providecommand\natexlab[1]{#1}
\providecommand\showeprint[2][]{arXiv:#2}

\bibitem[\protect\citeauthoryear{??}{pow}{[n.d.]}]%
        {powerpoint}
 \bibinfo{year}{[n.d.]}\natexlab{}.
\newblock \bibinfo{booktitle}{\emph{Create professional slide layouts with
  PowerPoint Designer}}.
\newblock
\urldef\tempurl%
\url{https://support.microsoft.com/en-us/office/create-professional-slide-layouts-with-powerpoint-designer-53c77d7b-dc40-45c2-b684-81415eac0617}
\showURL{%
\tempurl}


\bibitem[\protect\citeauthoryear{Ackerman and Palen}{Ackerman and
  Palen}{1996}]%
        {ackerman1996zephyr}
\bibfield{author}{\bibinfo{person}{Mark~S Ackerman} {and}
  \bibinfo{person}{Leysia Palen}.} \bibinfo{year}{1996}\natexlab{}.
\newblock \showarticletitle{The Zephyr Help Instance: promoting ongoing
  activity in a CSCW system}. In \bibinfo{booktitle}{\emph{Proceedings of the
  SIGCHI Conference on Human Factors in Computing Systems}}.
  \bibinfo{pages}{268--275}.
\newblock


\bibitem[\protect\citeauthoryear{Ai, Dumais, Craswell, and Liebling}{Ai
  et~al\mbox{.}}{2017}]%
        {ai2017characterizing}
\bibfield{author}{\bibinfo{person}{Qingyao Ai}, \bibinfo{person}{Susan~T
  Dumais}, \bibinfo{person}{Nick Craswell}, {and} \bibinfo{person}{Dan
  Liebling}.} \bibinfo{year}{2017}\natexlab{}.
\newblock \showarticletitle{Characterizing email search using large-scale
  behavioral logs and surveys}. In \bibinfo{booktitle}{\emph{Proceedings of the
  26th International Conference on World Wide Web}}.
  \bibinfo{pages}{1511--1520}.
\newblock


\bibitem[\protect\citeauthoryear{Bernstein, Little, Miller, Hartmann, Ackerman,
  Karger, Crowell, and Panovich}{Bernstein et~al\mbox{.}}{2010}]%
        {bernstein2010soylent}
\bibfield{author}{\bibinfo{person}{Michael~S Bernstein}, \bibinfo{person}{Greg
  Little}, \bibinfo{person}{Robert~C Miller}, \bibinfo{person}{Bj{\"o}rn
  Hartmann}, \bibinfo{person}{Mark~S Ackerman}, \bibinfo{person}{David~R
  Karger}, \bibinfo{person}{David Crowell}, {and} \bibinfo{person}{Katrina
  Panovich}.} \bibinfo{year}{2010}\natexlab{}.
\newblock \showarticletitle{Soylent: a word processor with a crowd inside}. In
  \bibinfo{booktitle}{\emph{Proceedings of the 23nd annual ACM symposium on
  User interface software and technology}}. \bibinfo{pages}{313--322}.
\newblock


\bibitem[\protect\citeauthoryear{Birnholtz and Ibara}{Birnholtz and
  Ibara}{2012}]%
        {birnholtz2012tracking}
\bibfield{author}{\bibinfo{person}{Jeremy Birnholtz} {and}
  \bibinfo{person}{Steven Ibara}.} \bibinfo{year}{2012}\natexlab{}.
\newblock \showarticletitle{Tracking changes in collaborative writing: edits,
  visibility and group maintenance}. In \bibinfo{booktitle}{\emph{Proceedings
  of the ACM 2012 conference on Computer Supported Cooperative Work}}.
  \bibinfo{pages}{809--818}.
\newblock


\bibitem[\protect\citeauthoryear{Birnholtz, Steinhardt, and Pavese}{Birnholtz
  et~al\mbox{.}}{2013}]%
        {birnholtz2013write}
\bibfield{author}{\bibinfo{person}{Jeremy Birnholtz},
  \bibinfo{person}{Stephanie Steinhardt}, {and} \bibinfo{person}{Antonella
  Pavese}.} \bibinfo{year}{2013}\natexlab{}.
\newblock \showarticletitle{Write here, write now! An experimental study of
  group maintenance in collaborative writing}. In
  \bibinfo{booktitle}{\emph{Proceedings of the SIGCHI Conference on Human
  Factors in Computing Systems}}. \bibinfo{pages}{961--970}.
\newblock


\bibitem[\protect\citeauthoryear{Bota, Fourney, Dumais, Religa, and
  Rounthwaite}{Bota et~al\mbox{.}}{2018}]%
        {bota2018characterizing}
\bibfield{author}{\bibinfo{person}{Horatiu Bota}, \bibinfo{person}{Adam
  Fourney}, \bibinfo{person}{Susan~T Dumais}, \bibinfo{person}{Tomasz~L
  Religa}, {and} \bibinfo{person}{Robert Rounthwaite}.}
  \bibinfo{year}{2018}\natexlab{}.
\newblock \showarticletitle{Characterizing Search Behavior in Productivity
  Software}. In \bibinfo{booktitle}{\emph{Proceedings of the 2018 Conference on
  Human Information Interaction \& Retrieval}}. \bibinfo{pages}{160--169}.
\newblock


\bibitem[\protect\citeauthoryear{Chen, Fisch, Weston, and Bordes}{Chen
  et~al\mbox{.}}{2017}]%
        {chen2017reading}
\bibfield{author}{\bibinfo{person}{Danqi Chen}, \bibinfo{person}{Adam Fisch},
  \bibinfo{person}{Jason Weston}, {and} \bibinfo{person}{Antoine Bordes}.}
  \bibinfo{year}{2017}\natexlab{}.
\newblock \showarticletitle{Reading wikipedia to answer open-domain questions}.
\newblock \bibinfo{journal}{\emph{arXiv preprint arXiv:1704.00051}}
  (\bibinfo{year}{2017}).
\newblock


\bibitem[\protect\citeauthoryear{Chilana, Ko, and Wobbrock}{Chilana
  et~al\mbox{.}}{2012}]%
        {chilana2012lemonaid}
\bibfield{author}{\bibinfo{person}{Parmit~K Chilana}, \bibinfo{person}{Andrew~J
  Ko}, {and} \bibinfo{person}{Jacob~O Wobbrock}.}
  \bibinfo{year}{2012}\natexlab{}.
\newblock \showarticletitle{LemonAid: selection-based crowdsourced contextual
  help for web applications}. In \bibinfo{booktitle}{\emph{Proceedings of the
  SIGCHI Conference on Human Factors in Computing Systems}}.
  \bibinfo{pages}{1549--1558}.
\newblock


\bibitem[\protect\citeauthoryear{Cranshaw, Elwany, Newman, Kocielnik, Yu, Soni,
  Teevan, and Monroy-Hern{\'a}ndez}{Cranshaw et~al\mbox{.}}{2017}]%
        {cranshaw2017calendar}
\bibfield{author}{\bibinfo{person}{Justin Cranshaw}, \bibinfo{person}{Emad
  Elwany}, \bibinfo{person}{Todd Newman}, \bibinfo{person}{Rafal Kocielnik},
  \bibinfo{person}{Bowen Yu}, \bibinfo{person}{Sandeep Soni},
  \bibinfo{person}{Jaime Teevan}, {and} \bibinfo{person}{Andr{\'e}s
  Monroy-Hern{\'a}ndez}.} \bibinfo{year}{2017}\natexlab{}.
\newblock \showarticletitle{Calendar. help: Designing a workflow-based
  scheduling agent with humans in the loop}. In
  \bibinfo{booktitle}{\emph{Proceedings of the 2017 CHI Conference on Human
  Factors in Computing Systems}}. \bibinfo{pages}{2382--2393}.
\newblock


\bibitem[\protect\citeauthoryear{Dredze}{Dredze}{2009}]%
        {dredze2009intelligent}
\bibfield{author}{\bibinfo{person}{Mark Dredze}.}
  \bibinfo{year}{2009}\natexlab{}.
\newblock \showarticletitle{Intelligent email: Aiding users with AI}.
\newblock \bibinfo{journal}{\emph{University of Pennsylvania, Philadelphia,
  PA}} (\bibinfo{year}{2009}).
\newblock


\bibitem[\protect\citeauthoryear{Dunn, Sagun, Higgins, Guney, Cirik, and
  Cho}{Dunn et~al\mbox{.}}{2017}]%
        {dunn2017searchqa}
\bibfield{author}{\bibinfo{person}{Matthew Dunn}, \bibinfo{person}{Levent
  Sagun}, \bibinfo{person}{Mike Higgins}, \bibinfo{person}{V~Ugur Guney},
  \bibinfo{person}{Volkan Cirik}, {and} \bibinfo{person}{Kyunghyun Cho}.}
  \bibinfo{year}{2017}\natexlab{}.
\newblock \showarticletitle{Searchqa: A new q\&a dataset augmented with context
  from a search engine}.
\newblock \bibinfo{journal}{\emph{arXiv preprint arXiv:1704.05179}}
  (\bibinfo{year}{2017}).
\newblock


\bibitem[\protect\citeauthoryear{Fast, Chen, Mendelsohn, Bassen, and
  Bernstein}{Fast et~al\mbox{.}}{2018}]%
        {fast2018iris}
\bibfield{author}{\bibinfo{person}{Ethan Fast}, \bibinfo{person}{Binbin Chen},
  \bibinfo{person}{Julia Mendelsohn}, \bibinfo{person}{Jonathan Bassen}, {and}
  \bibinfo{person}{Michael~S Bernstein}.} \bibinfo{year}{2018}\natexlab{}.
\newblock \showarticletitle{Iris: A conversational agent for complex tasks}. In
  \bibinfo{booktitle}{\emph{Proceedings of the 2018 CHI Conference on Human
  Factors in Computing Systems}}. \bibinfo{pages}{1--12}.
\newblock


\bibitem[\protect\citeauthoryear{Ferguson, Allen, Galescu, Quinn, and
  Swift}{Ferguson et~al\mbox{.}}{2009}]%
        {ferguson2009cardiac}
\bibfield{author}{\bibinfo{person}{George Ferguson}, \bibinfo{person}{James
  Allen}, \bibinfo{person}{Lucian Galescu}, \bibinfo{person}{Jill Quinn}, {and}
  \bibinfo{person}{Mary Swift}.} \bibinfo{year}{2009}\natexlab{}.
\newblock \showarticletitle{Cardiac: An intelligent conversational assistant
  for chronic heart failure patient heath monitoring}. In
  \bibinfo{booktitle}{\emph{2009 AAAI Fall Symposium Series}}.
\newblock


\bibitem[\protect\citeauthoryear{Gao and Cucerzan}{Gao and Cucerzan}{2017}]%
        {gao2017entity}
\bibfield{author}{\bibinfo{person}{Ning Gao} {and} \bibinfo{person}{Silviu
  Cucerzan}.} \bibinfo{year}{2017}\natexlab{}.
\newblock \showarticletitle{Entity linking to one thousand knowledge bases}. In
  \bibinfo{booktitle}{\emph{European Conference on Information Retrieval}}.
  Springer, \bibinfo{pages}{1--14}.
\newblock


\bibitem[\protect\citeauthoryear{Gelbukh, Sidorov, and Guzman-Arenas}{Gelbukh
  et~al\mbox{.}}{1999}]%
        {gelbukh1999document}
\bibfield{author}{\bibinfo{person}{Alexander Gelbukh}, \bibinfo{person}{Grigori
  Sidorov}, {and} \bibinfo{person}{Adolfo Guzman-Arenas}.}
  \bibinfo{year}{1999}\natexlab{}.
\newblock \showarticletitle{Document comparison with a weighted topic
  hierarchy}. In \bibinfo{booktitle}{\emph{Proceedings. Tenth International
  Workshop on Database and Expert Systems Applications. DEXA 99}}. IEEE,
  \bibinfo{pages}{566--570}.
\newblock


\bibitem[\protect\citeauthoryear{Hallett}{Hallett}{2008}]%
        {hallett2008multi}
\bibfield{author}{\bibinfo{person}{Catalina Hallett}.}
  \bibinfo{year}{2008}\natexlab{}.
\newblock \showarticletitle{Multi-modal presentation of medical histories}. In
  \bibinfo{booktitle}{\emph{Proceedings of the 13th international conference on
  Intelligent user interfaces}}. \bibinfo{pages}{80--89}.
\newblock


\bibitem[\protect\citeauthoryear{Hallett, Power, and Scott}{Hallett
  et~al\mbox{.}}{2006}]%
        {hallett2006summarisation}
\bibfield{author}{\bibinfo{person}{Catalina Hallett}, \bibinfo{person}{Richard
  Power}, {and} \bibinfo{person}{Donia Scott}.}
  \bibinfo{year}{2006}\natexlab{}.
\newblock \showarticletitle{Summarisation and visualisation of e-health data
  repositories}.
\newblock  (\bibinfo{year}{2006}).
\newblock


\bibitem[\protect\citeauthoryear{Han, Wang, Bendersky, and Najork}{Han
  et~al\mbox{.}}{2020}]%
        {han2020learning}
\bibfield{author}{\bibinfo{person}{Shuguang Han}, \bibinfo{person}{Xuanhui
  Wang}, \bibinfo{person}{Mike Bendersky}, {and} \bibinfo{person}{Marc
  Najork}.} \bibinfo{year}{2020}\natexlab{}.
\newblock \showarticletitle{Learning-to-Rank with BERT in TF-Ranking}.
\newblock \bibinfo{journal}{\emph{arXiv preprint arXiv:2004.08476}}
  (\bibinfo{year}{2020}).
\newblock


\bibitem[\protect\citeauthoryear{Hempel}{Hempel}{2015}]%
        {hempel2015facebook}
\bibfield{author}{\bibinfo{person}{Jessi Hempel}.}
  \bibinfo{year}{2015}\natexlab{}.
\newblock \showarticletitle{Facebook launches M, its bold answer to Siri and
  Cortana}.
\newblock \bibinfo{journal}{\emph{Wired. Retrieved January}}
  \bibinfo{volume}{1} (\bibinfo{year}{2015}), \bibinfo{pages}{2017}.
\newblock


\bibitem[\protect\citeauthoryear{Horowitz and Kamvar}{Horowitz and
  Kamvar}{2010}]%
        {horowitz2010anatomy}
\bibfield{author}{\bibinfo{person}{Damon Horowitz} {and}
  \bibinfo{person}{Sepandar~D Kamvar}.} \bibinfo{year}{2010}\natexlab{}.
\newblock \showarticletitle{The anatomy of a large-scale social search engine}.
  In \bibinfo{booktitle}{\emph{Proceedings of the 19th international conference
  on World wide web}}. \bibinfo{pages}{431--440}.
\newblock


\bibitem[\protect\citeauthoryear{Hoy}{Hoy}{2018}]%
        {hoy2018alexa}
\bibfield{author}{\bibinfo{person}{Matthew~B Hoy}.}
  \bibinfo{year}{2018}\natexlab{}.
\newblock \showarticletitle{Alexa, Siri, Cortana, and more: an introduction to
  voice assistants}.
\newblock \bibinfo{journal}{\emph{Medical reference services quarterly}}
  \bibinfo{volume}{37}, \bibinfo{number}{1} (\bibinfo{year}{2018}),
  \bibinfo{pages}{81--88}.
\newblock


\bibitem[\protect\citeauthoryear{Huet}{Huet}{2016}]%
        {huet2016humans}
\bibfield{author}{\bibinfo{person}{Ellen Huet}.}
  \bibinfo{year}{2016}\natexlab{}.
\newblock \showarticletitle{The Humans hiding behind the chatbots}.
\newblock \bibinfo{journal}{\emph{Bloomberg. com (April, 18, 2016) https://www.
  cnet. com/news/facebook-is-killing-m-its-personal-chatbot-assistant}}
  (\bibinfo{year}{2016}).
\newblock


\bibitem[\protect\citeauthoryear{Hutchinson, Mackay, Westerlund, Bederson,
  Druin, Plaisant, Beaudouin-Lafon, Conversy, Evans, Hansen,
  et~al\mbox{.}}{Hutchinson et~al\mbox{.}}{2003}]%
        {hutchinson2003technology}
\bibfield{author}{\bibinfo{person}{Hilary Hutchinson}, \bibinfo{person}{Wendy
  Mackay}, \bibinfo{person}{Bo Westerlund}, \bibinfo{person}{Benjamin~B
  Bederson}, \bibinfo{person}{Allison Druin}, \bibinfo{person}{Catherine
  Plaisant}, \bibinfo{person}{Michel Beaudouin-Lafon},
  \bibinfo{person}{St{\'e}phane Conversy}, \bibinfo{person}{Helen Evans},
  \bibinfo{person}{Heiko Hansen}, {et~al\mbox{.}}}
  \bibinfo{year}{2003}\natexlab{}.
\newblock \showarticletitle{Technology probes: inspiring design for and with
  families}. In \bibinfo{booktitle}{\emph{Proceedings of the SIGCHI conference
  on Human factors in computing systems}}. \bibinfo{pages}{17--24}.
\newblock


\bibitem[\protect\citeauthoryear{Iqbal, Teevan, Liebling, and Thompson}{Iqbal
  et~al\mbox{.}}{2018}]%
        {iqbal2018multitasking}
\bibfield{author}{\bibinfo{person}{Shamsi~T Iqbal}, \bibinfo{person}{Jaime
  Teevan}, \bibinfo{person}{Dan Liebling}, {and} \bibinfo{person}{Anne~Loomis
  Thompson}.} \bibinfo{year}{2018}\natexlab{}.
\newblock \showarticletitle{Multitasking with Play Write, a mobile
  microproductivity writing tool}. In \bibinfo{booktitle}{\emph{Proceedings of
  the 31st Annual ACM Symposium on User Interface Software and Technology}}.
  \bibinfo{pages}{411--422}.
\newblock


\bibitem[\protect\citeauthoryear{Iyyer, Boyd-Graber, Claudino, Socher, and
  Daum{\'e}~III}{Iyyer et~al\mbox{.}}{2014}]%
        {iyyer2014neural}
\bibfield{author}{\bibinfo{person}{Mohit Iyyer}, \bibinfo{person}{Jordan
  Boyd-Graber}, \bibinfo{person}{Leonardo Claudino}, \bibinfo{person}{Richard
  Socher}, {and} \bibinfo{person}{Hal Daum{\'e}~III}.}
  \bibinfo{year}{2014}\natexlab{}.
\newblock \showarticletitle{A neural network for factoid question answering
  over paragraphs}. In \bibinfo{booktitle}{\emph{Proceedings of the 2014
  conference on empirical methods in natural language processing (EMNLP)}}.
  \bibinfo{pages}{633--644}.
\newblock


\bibitem[\protect\citeauthoryear{Jahanbakhsh, Awadallah, Dumais, and
  Xu}{Jahanbakhsh et~al\mbox{.}}{2020}]%
        {jahanbakhsh2020effects}
\bibfield{author}{\bibinfo{person}{Farnaz Jahanbakhsh},
  \bibinfo{person}{Ahmed~Hassan Awadallah}, \bibinfo{person}{Susan~T Dumais},
  {and} \bibinfo{person}{Xuhai Xu}.} \bibinfo{year}{2020}\natexlab{}.
\newblock \showarticletitle{Effects of Past Interactions on User Experience
  with Recommended Documents}. In \bibinfo{booktitle}{\emph{Proceedings of the
  2020 Conference on Human Information Interaction and Retrieval}}.
  \bibinfo{pages}{153--162}.
\newblock


\bibitem[\protect\citeauthoryear{Jared~Spataro}{Jared~Spataro}{[n.d.]}]%
        {Word2018todo}
\bibfield{author}{\bibinfo{person}{Corporate Vice President for Microsoft~365
  Jared~Spataro}.} \bibinfo{year}{[n.d.]}\natexlab{}.
\newblock \bibinfo{booktitle}{\emph{Collaborate with others and keep track of
  to-dos with new AI features in Word}}.
\newblock
\urldef\tempurl%
\url{https://www.microsoft.com/en-us/microsoft-365/blog/2018/11/07/collaborate-with-others-and-keep-track-of-to-dos-with-new-ai-features-in-word/}
\showURL{%
\tempurl}


\bibitem[\protect\citeauthoryear{Joshi, Choi, Weld, and Zettlemoyer}{Joshi
  et~al\mbox{.}}{2017}]%
        {joshi2017triviaqa}
\bibfield{author}{\bibinfo{person}{Mandar Joshi}, \bibinfo{person}{Eunsol
  Choi}, \bibinfo{person}{Daniel~S Weld}, {and} \bibinfo{person}{Luke
  Zettlemoyer}.} \bibinfo{year}{2017}\natexlab{}.
\newblock \showarticletitle{Triviaqa: A large scale distantly supervised
  challenge dataset for reading comprehension}.
\newblock \bibinfo{journal}{\emph{arXiv preprint arXiv:1705.03551}}
  (\bibinfo{year}{2017}).
\newblock


\bibitem[\protect\citeauthoryear{Kamar and Redmond}{Kamar and Redmond}{2016}]%
        {kamar2016hybrid}
\bibfield{author}{\bibinfo{person}{Ece Kamar} {and} \bibinfo{person}{WA
  Redmond}.} \bibinfo{year}{2016}\natexlab{}.
\newblock \showarticletitle{Hybrid Intelligence and the Future of Work}. In
  \bibinfo{booktitle}{\emph{Productivity Decomposed: Getting Big Things Done
  with Little Microtasks Workshop (CHI 2016). http://research. microsoft.
  com/en-us/um/people/eckamar/papers/HybridIntelligence. pdf}}.
\newblock


\bibitem[\protect\citeauthoryear{Kaur, Gordon, Yang, Bigham, Teevan, Kamar, and
  Lasecki}{Kaur et~al\mbox{.}}{2017}]%
        {kaur2017crowdmask}
\bibfield{author}{\bibinfo{person}{Harmanpreet Kaur},
  \bibinfo{person}{Mitchell~L Gordon}, \bibinfo{person}{Yi~Wei Yang},
  \bibinfo{person}{Jeffrey~P Bigham}, \bibinfo{person}{Jaime Teevan},
  \bibinfo{person}{Ece Kamar}, {and} \bibinfo{person}{Walter~S Lasecki}.}
  \bibinfo{year}{2017}\natexlab{}.
\newblock \showarticletitle{CrowdMask: Using Crowds to Preserve Privacy in
  Crowd-Powered Systems via Progressive Filtering.}. In
  \bibinfo{booktitle}{\emph{HCOMP}}. \bibinfo{pages}{89--97}.
\newblock


\bibitem[\protect\citeauthoryear{Kumaran and Carvalho}{Kumaran and
  Carvalho}{2009}]%
        {kumaran2009reducing}
\bibfield{author}{\bibinfo{person}{Giridhar Kumaran} {and}
  \bibinfo{person}{Vitor~R Carvalho}.} \bibinfo{year}{2009}\natexlab{}.
\newblock \showarticletitle{Reducing long queries using query quality
  predictors}. In \bibinfo{booktitle}{\emph{Proceedings of the 32nd
  international ACM SIGIR conference on Research and development in information
  retrieval}}. \bibinfo{pages}{564--571}.
\newblock


\bibitem[\protect\citeauthoryear{Kwiatkowski, Palomaki, Redfield, Collins,
  Parikh, Alberti, Epstein, Polosukhin, Devlin, Lee, et~al\mbox{.}}{Kwiatkowski
  et~al\mbox{.}}{2019}]%
        {kwiatkowski2019natural}
\bibfield{author}{\bibinfo{person}{Tom Kwiatkowski},
  \bibinfo{person}{Jennimaria Palomaki}, \bibinfo{person}{Olivia Redfield},
  \bibinfo{person}{Michael Collins}, \bibinfo{person}{Ankur Parikh},
  \bibinfo{person}{Chris Alberti}, \bibinfo{person}{Danielle Epstein},
  \bibinfo{person}{Illia Polosukhin}, \bibinfo{person}{Jacob Devlin},
  \bibinfo{person}{Kenton Lee}, {et~al\mbox{.}}}
  \bibinfo{year}{2019}\natexlab{}.
\newblock \showarticletitle{Natural questions: a benchmark for question
  answering research}.
\newblock \bibinfo{journal}{\emph{Transactions of the Association for
  Computational Linguistics}}  \bibinfo{volume}{7} (\bibinfo{year}{2019}),
  \bibinfo{pages}{453--466}.
\newblock


\bibitem[\protect\citeauthoryear{Landis and Koch}{Landis and Koch}{1977}]%
        {landis1977measurement}
\bibfield{author}{\bibinfo{person}{J~Richard Landis} {and}
  \bibinfo{person}{Gary~G Koch}.} \bibinfo{year}{1977}\natexlab{}.
\newblock \showarticletitle{The measurement of observer agreement for
  categorical data}.
\newblock \bibinfo{journal}{\emph{biometrics}} (\bibinfo{year}{1977}),
  \bibinfo{pages}{159--174}.
\newblock


\bibitem[\protect\citeauthoryear{Larson and Csikszentmihalyi}{Larson and
  Csikszentmihalyi}{2014}]%
        {larson2014experience}
\bibfield{author}{\bibinfo{person}{Reed Larson} {and} \bibinfo{person}{Mihaly
  Csikszentmihalyi}.} \bibinfo{year}{2014}\natexlab{}.
\newblock \showarticletitle{The experience sampling method}.
\newblock In \bibinfo{booktitle}{\emph{Flow and the foundations of positive
  psychology}}. \bibinfo{publisher}{Springer}, \bibinfo{pages}{21--34}.
\newblock


\bibitem[\protect\citeauthoryear{Lasecki, Gordon, Teevan, Kamar, and
  Bigham}{Lasecki et~al\mbox{.}}{2015}]%
        {lasecki2015preserving}
\bibfield{author}{\bibinfo{person}{Walter~S Lasecki}, \bibinfo{person}{Mitchell
  Gordon}, \bibinfo{person}{Jaime Teevan}, \bibinfo{person}{Ece Kamar}, {and}
  \bibinfo{person}{Jeffrey~P Bigham}.} \bibinfo{year}{2015}\natexlab{}.
\newblock \showarticletitle{Preserving privacy in crowd-powered systems}. In
  \bibinfo{booktitle}{\emph{Proceedings of AAMAS 2015 Workshop on Human-Agent
  Interaction Design and Models}}.
\newblock


\bibitem[\protect\citeauthoryear{Lasecki, Wesley, Nichols, Kulkarni, Allen, and
  Bigham}{Lasecki et~al\mbox{.}}{2013}]%
        {lasecki2013chorus}
\bibfield{author}{\bibinfo{person}{Walter~S Lasecki}, \bibinfo{person}{Rachel
  Wesley}, \bibinfo{person}{Jeffrey Nichols}, \bibinfo{person}{Anand Kulkarni},
  \bibinfo{person}{James~F Allen}, {and} \bibinfo{person}{Jeffrey~P Bigham}.}
  \bibinfo{year}{2013}\natexlab{}.
\newblock \showarticletitle{Chorus: a crowd-powered conversational assistant}.
  In \bibinfo{booktitle}{\emph{Proceedings of the 26th annual ACM symposium on
  User interface software and technology}}. \bibinfo{pages}{151--162}.
\newblock


\bibitem[\protect\citeauthoryear{Martelaro, Teevan, and Iqbal}{Martelaro
  et~al\mbox{.}}{2019}]%
        {martelaro2019}
\bibfield{author}{\bibinfo{person}{Nikolas Martelaro}, \bibinfo{person}{Jaime
  Teevan}, {and} \bibinfo{person}{Shamsi~T. Iqbal}.}
  \bibinfo{year}{2019}\natexlab{}.
\newblock \showarticletitle{An Exploration of Speech-Based Productivity Support
  in the Car}. In \bibinfo{booktitle}{\emph{Proceedings of the 2019 CHI
  Conference on Human Factors in Computing Systems}} (Glasgow, Scotland Uk)
  \emph{(\bibinfo{series}{CHI '19})}. \bibinfo{publisher}{Association for
  Computing Machinery}, \bibinfo{address}{New York, NY, USA},
  \bibinfo{pages}{1–12}.
\newblock
\showISBNx{9781450359702}
\urldef\tempurl%
\url{https://doi.org/10.1145/3290605.3300494}
\showDOI{\tempurl}


\bibitem[\protect\citeauthoryear{Metz}{Metz}{2015}]%
        {cade2015wired}
\bibfield{author}{\bibinfo{person}{Cade Metz}.}
  \bibinfo{year}{2015}\natexlab{}.
\newblock \showarticletitle{AI helps humans best when humans help the AI.}
\newblock \bibinfo{journal}{\emph{Wired.com (2015)
  https://www.wired.com/2015/09/ai-helps-humans-best-humans-help-ai}}
  \bibinfo{volume}{1} (\bibinfo{year}{2015}).
\newblock


\bibitem[\protect\citeauthoryear{Murnane}{Murnane}{[n.d.]}]%
        {IFTT2017Forbes}
\bibfield{author}{\bibinfo{person}{Kevin Murnane}.}
  \bibinfo{year}{[n.d.]}\natexlab{}.
\newblock \bibinfo{booktitle}{\emph{IFTTT Survey Provides Insight Into What
  People Do With Amazon's Echo And Google's Home}}.
\newblock
\urldef\tempurl%
\url{https://www.forbes.com/sites/kevinmurnane/2017/07/12/ifttt-survey-provides-insight-into-what-people-do-with-voice-controlled-assistants/#2d5966b643e6}
\showURL{%
\tempurl}


\bibitem[\protect\citeauthoryear{Murray, Gopinath, Agrawal, Horng, Sontag, and
  Karger}{Murray et~al\mbox{.}}{2021}]%
        {Murray2021MedKnowts}
\bibfield{author}{\bibinfo{person}{Luke Murray}, \bibinfo{person}{Divya
  Gopinath}, \bibinfo{person}{Monica Agrawal}, \bibinfo{person}{Steven Horng},
  \bibinfo{person}{David Sontag}, {and} \bibinfo{person}{David~R. Karger}.}
  \bibinfo{year}{2021}\natexlab{}.
\newblock \showarticletitle{MedKnowts: Unified Documentation and Information
  Retrieval for Electronic Health Records}. In \bibinfo{booktitle}{\emph{The
  34th Annual ACM Symposium on User Interface Software and Technology (UIST
  ’21)}}.
\newblock


\bibitem[\protect\citeauthoryear{Myers, Berry, Blythe, Conley, Gervasio,
  McGuinness, Morley, Pfeffer, Pollack, and Tambe}{Myers et~al\mbox{.}}{2007}]%
        {myers2007intelligent}
\bibfield{author}{\bibinfo{person}{Karen Myers}, \bibinfo{person}{Pauline
  Berry}, \bibinfo{person}{Jim Blythe}, \bibinfo{person}{Ken Conley},
  \bibinfo{person}{Melinda Gervasio}, \bibinfo{person}{Deborah~L McGuinness},
  \bibinfo{person}{David Morley}, \bibinfo{person}{Avi Pfeffer},
  \bibinfo{person}{Martha Pollack}, {and} \bibinfo{person}{Milind Tambe}.}
  \bibinfo{year}{2007}\natexlab{}.
\newblock \showarticletitle{An intelligent personal assistant for task and time
  management}.
\newblock \bibinfo{journal}{\emph{AI Magazine}} \bibinfo{volume}{28},
  \bibinfo{number}{2} (\bibinfo{year}{2007}), \bibinfo{pages}{47--47}.
\newblock


\bibitem[\protect\citeauthoryear{Nguyen, Rosenberg, Song, Gao, Tiwary,
  Majumder, and Deng}{Nguyen et~al\mbox{.}}{2016}]%
        {nguyen2016ms}
\bibfield{author}{\bibinfo{person}{Tri Nguyen}, \bibinfo{person}{Mir
  Rosenberg}, \bibinfo{person}{Xia Song}, \bibinfo{person}{Jianfeng Gao},
  \bibinfo{person}{Saurabh Tiwary}, \bibinfo{person}{Rangan Majumder}, {and}
  \bibinfo{person}{Li Deng}.} \bibinfo{year}{2016}\natexlab{}.
\newblock \showarticletitle{Ms marco: A human-generated machine reading
  comprehension dataset}.
\newblock  (\bibinfo{year}{2016}).
\newblock


\bibitem[\protect\citeauthoryear{Olson and Kemery}{Olson and Kemery}{2019}]%
        {olson2019voice}
\bibfield{author}{\bibinfo{person}{Christi Olson} {and} \bibinfo{person}{Kelli
  Kemery}.} \bibinfo{year}{2019}\natexlab{}.
\newblock \showarticletitle{Voice report: From answers to action: customer
  adoption of voice technology and digital assistants}.
\newblock \bibinfo{journal}{\emph{Microsoft Search and Market Intelligence,
  Tech. Rep}} (\bibinfo{year}{2019}).
\newblock


\bibitem[\protect\citeauthoryear{Olson, Wang, Olson, and Zhang}{Olson
  et~al\mbox{.}}{2017}]%
        {olson2017people}
\bibfield{author}{\bibinfo{person}{Judith~S Olson}, \bibinfo{person}{Dakuo
  Wang}, \bibinfo{person}{Gary~M Olson}, {and} \bibinfo{person}{Jingwen
  Zhang}.} \bibinfo{year}{2017}\natexlab{}.
\newblock \showarticletitle{How people write together now: Beginning the
  investigation with advanced undergraduates in a project course}.
\newblock \bibinfo{journal}{\emph{ACM Transactions on Computer-Human
  Interaction (TOCHI)}} \bibinfo{volume}{24}, \bibinfo{number}{1}
  (\bibinfo{year}{2017}), \bibinfo{pages}{1--40}.
\newblock


\bibitem[\protect\citeauthoryear{Persky}{Persky}{[n.d.]}]%
        {Cooking2017GoogleHome}
\bibfield{author}{\bibinfo{person}{Emma Persky}.}
  \bibinfo{year}{[n.d.]}\natexlab{}.
\newblock \bibinfo{booktitle}{\emph{Now we're cooking – the Assistant on
  Google Home is your secret ingredient}}.
\newblock
\urldef\tempurl%
\url{https://www.blog.google/products/assistant/cooking-with-the-assistant-google-home-your-secret-ingredient/}
\showURL{%
\tempurl}


\bibitem[\protect\citeauthoryear{Posner and Baecker}{Posner and
  Baecker}{1992}]%
        {posner1992people}
\bibfield{author}{\bibinfo{person}{Ilona~R Posner} {and}
  \bibinfo{person}{Ronald~M Baecker}.} \bibinfo{year}{1992}\natexlab{}.
\newblock \showarticletitle{How people write together (groupware)}. In
  \bibinfo{booktitle}{\emph{Proceedings of the Twenty-Fifth Hawaii
  International Conference on System Sciences}}, Vol.~\bibinfo{volume}{4}.
  IEEE, \bibinfo{pages}{127--138}.
\newblock


\bibitem[\protect\citeauthoryear{Rajpurkar, Jia, and Liang}{Rajpurkar
  et~al\mbox{.}}{2018}]%
        {rajpurkar2018know}
\bibfield{author}{\bibinfo{person}{Pranav Rajpurkar}, \bibinfo{person}{Robin
  Jia}, {and} \bibinfo{person}{Percy Liang}.} \bibinfo{year}{2018}\natexlab{}.
\newblock \showarticletitle{Know what you don't know: Unanswerable questions
  for SQuAD}.
\newblock \bibinfo{journal}{\emph{arXiv preprint arXiv:1806.03822}}
  (\bibinfo{year}{2018}).
\newblock


\bibitem[\protect\citeauthoryear{Rajpurkar, Zhang, Lopyrev, and
  Liang}{Rajpurkar et~al\mbox{.}}{2016}]%
        {rajpurkar2016squad}
\bibfield{author}{\bibinfo{person}{Pranav Rajpurkar}, \bibinfo{person}{Jian
  Zhang}, \bibinfo{person}{Konstantin Lopyrev}, {and} \bibinfo{person}{Percy
  Liang}.} \bibinfo{year}{2016}\natexlab{}.
\newblock \showarticletitle{Squad: 100,000+ questions for machine comprehension
  of text}.
\newblock \bibinfo{journal}{\emph{arXiv preprint arXiv:1606.05250}}
  (\bibinfo{year}{2016}).
\newblock


\bibitem[\protect\citeauthoryear{Retelny, Robaszkiewicz, To, Lasecki, Patel,
  Rahmati, Doshi, Valentine, and Bernstein}{Retelny et~al\mbox{.}}{2014}]%
        {retelny2014expert}
\bibfield{author}{\bibinfo{person}{Daniela Retelny},
  \bibinfo{person}{S{\'e}bastien Robaszkiewicz}, \bibinfo{person}{Alexandra
  To}, \bibinfo{person}{Walter~S Lasecki}, \bibinfo{person}{Jay Patel},
  \bibinfo{person}{Negar Rahmati}, \bibinfo{person}{Tulsee Doshi},
  \bibinfo{person}{Melissa Valentine}, {and} \bibinfo{person}{Michael~S
  Bernstein}.} \bibinfo{year}{2014}\natexlab{}.
\newblock \showarticletitle{Expert crowdsourcing with flash teams}. In
  \bibinfo{booktitle}{\emph{Proceedings of the 27th annual ACM symposium on
  User interface software and technology}}. \bibinfo{pages}{75--85}.
\newblock


\bibitem[\protect\citeauthoryear{Richardson and White}{Richardson and
  White}{2011}]%
        {richardson2011supporting}
\bibfield{author}{\bibinfo{person}{Matthew Richardson} {and}
  \bibinfo{person}{Ryen~W White}.} \bibinfo{year}{2011}\natexlab{}.
\newblock \showarticletitle{Supporting synchronous social q\&a throughout the
  question lifecycle}. In \bibinfo{booktitle}{\emph{Proceedings of the 20th
  international conference on World wide web}}. \bibinfo{pages}{755--764}.
\newblock


\bibitem[\protect\citeauthoryear{Su, Awadallah, Khabsa, Pantel, Gamon, and
  Encarnacion}{Su et~al\mbox{.}}{2017}]%
        {su2017building}
\bibfield{author}{\bibinfo{person}{Yu Su}, \bibinfo{person}{Ahmed~Hassan
  Awadallah}, \bibinfo{person}{Madian Khabsa}, \bibinfo{person}{Patrick
  Pantel}, \bibinfo{person}{Michael Gamon}, {and} \bibinfo{person}{Mark
  Encarnacion}.} \bibinfo{year}{2017}\natexlab{}.
\newblock \showarticletitle{Building natural language interfaces to web apis}.
  In \bibinfo{booktitle}{\emph{Proceedings of the 2017 ACM on Conference on
  Information and Knowledge Management}}. \bibinfo{pages}{177--186}.
\newblock


\bibitem[\protect\citeauthoryear{Sultanum, Singh, Brudno, and
  Chevalier}{Sultanum et~al\mbox{.}}{2018}]%
        {sultanum2018doccurate}
\bibfield{author}{\bibinfo{person}{Nicole Sultanum}, \bibinfo{person}{Devin
  Singh}, \bibinfo{person}{Michael Brudno}, {and} \bibinfo{person}{Fanny
  Chevalier}.} \bibinfo{year}{2018}\natexlab{}.
\newblock \showarticletitle{Doccurate: A curation-based approach for clinical
  text visualization}.
\newblock \bibinfo{journal}{\emph{IEEE transactions on visualization and
  computer graphics}} \bibinfo{volume}{25}, \bibinfo{number}{1}
  (\bibinfo{year}{2018}), \bibinfo{pages}{142--151}.
\newblock


\bibitem[\protect\citeauthoryear{Sun, Lambert, Uchida, and Remy}{Sun
  et~al\mbox{.}}{2014}]%
        {sun2014collaboration}
\bibfield{author}{\bibinfo{person}{Yunting Sun}, \bibinfo{person}{Diane
  Lambert}, \bibinfo{person}{Makoto Uchida}, {and} \bibinfo{person}{Nicolas
  Remy}.} \bibinfo{year}{2014}\natexlab{}.
\newblock \showarticletitle{Collaboration in the cloud at Google}. In
  \bibinfo{booktitle}{\emph{Proceedings of the 2014 ACM conference on Web
  science}}. \bibinfo{pages}{239--240}.
\newblock


\bibitem[\protect\citeauthoryear{ter Hoeve, Sim, Nouri, Fourney, de~Rijke, and
  White}{ter Hoeve et~al\mbox{.}}{2020}]%
        {ter2020conversations}
\bibfield{author}{\bibinfo{person}{Maartje ter Hoeve}, \bibinfo{person}{Robert
  Sim}, \bibinfo{person}{Elnaz Nouri}, \bibinfo{person}{Adam Fourney},
  \bibinfo{person}{Maarten de Rijke}, {and} \bibinfo{person}{Ryen~W White}.}
  \bibinfo{year}{2020}\natexlab{}.
\newblock \showarticletitle{Conversations with Documents: An Exploration of
  Document-Centered Assistance}. In \bibinfo{booktitle}{\emph{Proceedings of
  the 2020 Conference on Human Information Interaction and Retrieval}}.
  \bibinfo{pages}{43--52}.
\newblock


\bibitem[\protect\citeauthoryear{Todi, Leiva, Buschek, Tian, and
  Oulasvirta}{Todi et~al\mbox{.}}{2021}]%
        {todi2021conversations}
\bibfield{author}{\bibinfo{person}{Kashyap Todi}, \bibinfo{person}{Luis~A
  Leiva}, \bibinfo{person}{Daniel Buschek}, \bibinfo{person}{Pin Tian}, {and}
  \bibinfo{person}{Antti Oulasvirta}.} \bibinfo{year}{2021}\natexlab{}.
\newblock \showarticletitle{Conversations with GUIs}. In
  \bibinfo{booktitle}{\emph{Designing Interactive Systems Conference 2021}}.
  \bibinfo{pages}{1447--1457}.
\newblock


\bibitem[\protect\citeauthoryear{Tur, Deoras, and Hakkani-T{\"u}r}{Tur
  et~al\mbox{.}}{2014}]%
        {tur2014detecting}
\bibfield{author}{\bibinfo{person}{Gokhan Tur}, \bibinfo{person}{Anoop Deoras},
  {and} \bibinfo{person}{Dilek Hakkani-T{\"u}r}.}
  \bibinfo{year}{2014}\natexlab{}.
\newblock \showarticletitle{Detecting out-of-domain utterances addressed to a
  virtual personal assistant}. In \bibinfo{booktitle}{\emph{Fifteenth Annual
  Conference of the International Speech Communication Association}}.
\newblock


\bibitem[\protect\citeauthoryear{Valentine, Retelny, To, Rahmati, Doshi, and
  Bernstein}{Valentine et~al\mbox{.}}{2017}]%
        {valentine2017flash}
\bibfield{author}{\bibinfo{person}{Melissa~A Valentine},
  \bibinfo{person}{Daniela Retelny}, \bibinfo{person}{Alexandra To},
  \bibinfo{person}{Negar Rahmati}, \bibinfo{person}{Tulsee Doshi}, {and}
  \bibinfo{person}{Michael~S Bernstein}.} \bibinfo{year}{2017}\natexlab{}.
\newblock \showarticletitle{Flash organizations: Crowdsourcing complex work by
  structuring crowds as organizations}. In
  \bibinfo{booktitle}{\emph{Proceedings of the 2017 CHI conference on human
  factors in computing systems}}. \bibinfo{pages}{3523--3537}.
\newblock


\bibitem[\protect\citeauthoryear{Wang, Tan, and Lu}{Wang et~al\mbox{.}}{2017}]%
        {wang2017users}
\bibfield{author}{\bibinfo{person}{Dakuo Wang}, \bibinfo{person}{Haodan Tan},
  {and} \bibinfo{person}{Tun Lu}.} \bibinfo{year}{2017}\natexlab{}.
\newblock \showarticletitle{Why users do not want to write together when they
  are writing together: Users' rationales for today's collaborative writing
  practices}.
\newblock \bibinfo{journal}{\emph{Proceedings of the ACM on Human-Computer
  Interaction}} \bibinfo{volume}{1}, \bibinfo{number}{CSCW}
  (\bibinfo{year}{2017}), \bibinfo{pages}{1--18}.
\newblock


\bibitem[\protect\citeauthoryear{White, Richardson, and Liu}{White
  et~al\mbox{.}}{2011}]%
        {white2011effects}
\bibfield{author}{\bibinfo{person}{Ryen~W White}, \bibinfo{person}{Matthew
  Richardson}, {and} \bibinfo{person}{Yandong Liu}.}
  \bibinfo{year}{2011}\natexlab{}.
\newblock \showarticletitle{Effects of community size and contact rate in
  synchronous social Q\&A}. In \bibinfo{booktitle}{\emph{Proceedings of the
  SIGCHI Conference on Human Factors in Computing Systems}}.
  \bibinfo{pages}{2837--2846}.
\newblock


\bibitem[\protect\citeauthoryear{Xiao, Wang, Yan, and Zheng}{Xiao
  et~al\mbox{.}}{2018}]%
        {xiao2018dual}
\bibfield{author}{\bibinfo{person}{Han Xiao}, \bibinfo{person}{Feng Wang},
  \bibinfo{person}{Jianfeng Yan}, {and} \bibinfo{person}{Jingyao Zheng}.}
  \bibinfo{year}{2018}\natexlab{}.
\newblock \showarticletitle{Dual ask-answer network for machine reading
  comprehension}.
\newblock \bibinfo{journal}{\emph{arXiv preprint arXiv:1809.01997}}
  (\bibinfo{year}{2018}).
\newblock


\bibitem[\protect\citeauthoryear{Yang, Yih, and Meek}{Yang
  et~al\mbox{.}}{2015}]%
        {yang2015wikiqa}
\bibfield{author}{\bibinfo{person}{Yi Yang}, \bibinfo{person}{Wen-tau Yih},
  {and} \bibinfo{person}{Christopher Meek}.} \bibinfo{year}{2015}\natexlab{}.
\newblock \showarticletitle{Wikiqa: A challenge dataset for open-domain
  question answering}. In \bibinfo{booktitle}{\emph{Proceedings of the 2015
  conference on empirical methods in natural language processing}}.
  \bibinfo{pages}{2013--2018}.
\newblock


\bibitem[\protect\citeauthoryear{Yang, Qi, Zhang, Bengio, Cohen, Salakhutdinov,
  and Manning}{Yang et~al\mbox{.}}{2018}]%
        {yang2018hotpotqa}
\bibfield{author}{\bibinfo{person}{Zhilin Yang}, \bibinfo{person}{Peng Qi},
  \bibinfo{person}{Saizheng Zhang}, \bibinfo{person}{Yoshua Bengio},
  \bibinfo{person}{William~W Cohen}, \bibinfo{person}{Ruslan Salakhutdinov},
  {and} \bibinfo{person}{Christopher~D Manning}.}
  \bibinfo{year}{2018}\natexlab{}.
\newblock \showarticletitle{Hotpotqa: A dataset for diverse, explainable
  multi-hop question answering}.
\newblock \bibinfo{journal}{\emph{arXiv preprint arXiv:1809.09600}}
  (\bibinfo{year}{2018}).
\newblock


\bibitem[\protect\citeauthoryear{Yao, Wan, and Xiao}{Yao et~al\mbox{.}}{2017}]%
        {yao2017recent}
\bibfield{author}{\bibinfo{person}{Jin-ge Yao}, \bibinfo{person}{Xiaojun Wan},
  {and} \bibinfo{person}{Jianguo Xiao}.} \bibinfo{year}{2017}\natexlab{}.
\newblock \showarticletitle{Recent advances in document summarization}.
\newblock \bibinfo{journal}{\emph{Knowledge and Information Systems}}
  \bibinfo{volume}{53}, \bibinfo{number}{2} (\bibinfo{year}{2017}),
  \bibinfo{pages}{297--336}.
\newblock


\bibitem[\protect\citeauthoryear{Yim, Wang, Olson, Vu, and Warschauer}{Yim
  et~al\mbox{.}}{2017}]%
        {yim2017synchronous}
\bibfield{author}{\bibinfo{person}{Soobin Yim}, \bibinfo{person}{Dakuo Wang},
  \bibinfo{person}{Judith Olson}, \bibinfo{person}{Viet Vu}, {and}
  \bibinfo{person}{Mark Warschauer}.} \bibinfo{year}{2017}\natexlab{}.
\newblock \showarticletitle{Synchronous writing in the classroom:
  Undergraduates’ collaborative practices and their impact on text quality,
  quantity, and style}. In \bibinfo{booktitle}{\emph{Proceedings of the
  Conference on Computer Supported Cooperative Work (CSCW’17)}},
  Vol.~\bibinfo{volume}{10}.
\newblock


\end{thebibliography}


\received{April 2021}
\received[revised]{November 2021}
\received[accepted]{March 2022}

\end{document}